\documentclass[12pt]{article}

\usepackage{amsmath,amssymb}           
\usepackage{amscd}                     
\usepackage{epsfig}                    
\usepackage[matrix,arrow]{xy}          
\usepackage{xspace}                    
\usepackage{stmaryrd}                  
\usepackage{slashed}
\usepackage{tikz}
\usepackage{array}

\usepackage{jheppub}                   

\DeclareSymbolFont{bbold}{U}{bbold}{m}{n}
\DeclareSymbolFontAlphabet{\mathbbold}{bbold}

\newcommand{\bq}{\begin{equation}}
\newcommand{\eq}{\end{equation}}
\newcommand{\bea}{\begin{eqnarray}}
\newcommand{\eea}{\end{eqnarray}}

\newcommand{\dd}{\mathrm{d}}
\newcommand{\ee}{\mathrm{e}}

\newcommand{\der}{\partial}

\newcommand{\bbR}{\mathbb{R}}

\DeclareMathOperator{\SU}{\mathit{SU}}
\DeclareMathOperator{\SO}{\mathit{SO}}
\DeclareMathOperator{\SL}{\mathit{SL}}
\DeclareMathOperator{\GL}{\mathit{GL}}

\DeclareMathOperator{\Spin}{\mathit{Spin}}

\newcommand{\rep}[1]{\mathbf{#1}}
\newcommand{\repp}[2]{(\rep{#1}, \rep{#2})}

\newcommand{\ph}[1]{\phantom{#1}}

\newcommand{\Lgen}{L}

\newcommand{\Dgen}{{D}}

\DeclareMathOperator{\adj}{ad}

\newcommand{\LC}{\nabla}

\newcommand{\proj}[1]{\times_{#1}}

\DeclareMathOperator{\Edd}{\mathit{E_{d(d)}}}
\DeclareMathOperator{\Hd}{\mathit{H_d}}

\newcommand{\tA}{{\tilde{A}}}
\newcommand{\tF}{{\tilde{F}}}

\newcommand{\am}{Q}

\newcommand{\Vout}{\curlywedge}

\newcommand{\DSS}{\slashed{\Dgen}}

\newcommand{\DSJ}{\Dgen \Vout}

\newcommand{\ra}{\rightarrow}
\newcommand{\lra}{\leftrightarrow}

\newcommand{\hE}{\hat{E}}

\newcommand{\Ehfd}{E^{(1/2)}_d}
\newcommand{\Hhfd}{H^{(1/2)}_d}

\title{Subsectors, Dynkin Diagrams and New Generalised Geometries} 

\author{
  Charles Strickland-Constable  
  }

\affiliation{II. Institut f\"ur Theoretische Physik 
   der Universit\"at Hamburg, \\
   Luruper Chaussee 149, D-22761 Hamburg, Germany \\}

\emailAdd{charles.strickland.constable@desy.de}

\subheader{\textrm{ZMP-HH/13-19}}

\abstract{
We examine how generalised geometries can be associated with a labelled Dynkin diagram built around a gravity line. We present a series of new generalised geometries based on the groups $\Spin(d,d)\times\bbR^+$ for which the generalised tangent space transforms in a spinor representation of the group. In low dimensions these all appear in subsectors of maximal supergravity theories. The case $d=8$ provides a geometry for eight-dimensional backgrounds of M theory with only seven-form flux, which have not been included in any previous geometric construction. This geometry is also one of a series of ``half-exceptional" geometries, which ``geometrise" a six-form gauge field. In the appendix, we consider examples of other algebras appearing in gravitational theories and give a method to derive the Dynkin labels for the ``section condition" in general. We argue that generalised geometry can describe restrictions and subsectors of many gravitational theories.
}

\begin{document}
\maketitle



\section{Introduction}
\label{sec:intro}


Generalised geometry~\cite{GCY,Gualtieri} is the study of structures, analogous to those of ordinary differential geometry, defined on an extended tangent space $E \simeq T \oplus \dots$, which is generically twisted by some gerbe (or ``gerbe-like") structure. In~\cite{CSW1,CSW2,CSW3}, it was shown that there is a very natural formulation of certain supergravity theories in the language of generalised geometry. This article serves as a discussion of how one might directly apply this construction to more general algebras and theories. Significant work in this direction~\cite{Baraglia} has already appeared in the mathematics literature\footnote{We thank Marco Gualtieri for pointing out the direct relevance of this reference to the research presented here.}, and here we will present some new examples.

Thus far, generalised geometries based around the groups $O(d,d)$ and $E_{d(d)}$ (for $d\leq7$) and their relevance to physics have been well-studied~\cite{CSW1,CSW2,CSW3,gen-susy1,gen-susy2,gen-susy3,gen-susy4,sigmas1,sigmas2,sigmas3,sigmas4,Gen-geom1,Gen-geom2,Gen-geom3,Gen-geom4,Gen-geom5,Gen-geom6,Gen-geom7,Gen-geom8,Gen-geom9,Gen-geom10,Gen-geom11,Gen-geom12,Gen-geom13,Gen-geom14,Gen-geom15,gen-calib1,gen-calib2,gen-calib3,gen-calib4,gen-calib5,Koerber:2010bx,Hull07,PW,Triendl:2009ap,GLSW,E7-flux,GO1,GO2,GT} (see also the literature on doubled constructions~\cite{siegel1,siegel2,Tfold,dft,Hohm:2010jy,DFT1,DFT2,DFT3,DFT3b,DFT4,DFT5,DFT6,DFT7,DFT8,DFT9,DFT10,DFT11,DFT12,DFT13} and other extended geometries~\cite{BP1,BP2,BP-alg,BPW,BCKT,MTheoryDFT1,MTheoryDFT2,MTheoryDFT3,MTheoryDFT4,MTheoryDFT5,MTheoryDFT6}). 
There has also been some work on generalised geometry for the groups $O(d,d+n)$~\cite{Baraglia,Andriot,BnGeom} (see also~\cite{Hohm:2011ex,Hohm:2011nu}) and recently reduction of Courant algebroids~\cite{courant-reduction} on principal bundles has been used to describe the non-abelian generalisation~\cite{heterotic1,heterotic2}.

However, as shown in~\cite{Baraglia}, one can associate similar Leibniz algebroids to more general classes of Lie algebras. In fact the only necessary condition is the existence of a $\GL(d,\bbR)$ subalgebra, under which the decomposition of the adjoint representation consists only of this subalgebra and exterior powers of the standard representation and its dual.

In this paper, we will endeavour to study more systematically the construction of other types of generalised geometry, but before we begin, we must explain in more detail what we mean by a generalised geometry. 
We will adopt a fairly conservative definition, requiring that the key attributes of the construction of~\cite{CSW1,CSW2,CSW3} hold good and taking inspiration from the above observation of~\cite{Baraglia}. 
We also keep the motivation of making contact with supergravity theories closely in mind. 
One could envisage a slightly more ambitious approach where the algebraic aspects of the decomposition of the generalised tangent space are defined simply by a section condition on it, discussed later in this paper. However, we leave this for the future as the definition we give here is adequate for our purposes and 
fits with the spirit of the remainder of the paper.

The key features we require of the generalised tangent space $E$ on a manifold, are as follows. Firstly, $E$ should be an extension (or sequence of extensions) of the usual tangent bundle $T$ by other ordinary $\GL(d,\bbR)$ tensor bundles, so that locally we have 
\begin{equation}
\label{eq:E-def}
	E \simeq T \oplus (\dots)
\end{equation}
where $(\dots)$ denotes the added tensor bundles. 
The fibre of $E$ should then naturally become a representation of some larger semi-simple structure group $G$ (often augmented by an additional $\bbR^+$ factor).  
The partial derivative of a function can then be thought of as living in the $T^*$ subbundle of $E^*$, 
which we require to be stabilised by a parabolic subgroup of $G$.
One can then write the general expression for the Dorfman derivative introduced in~\cite{CSW2}
\begin{equation*}
	\Lgen_V = \der_V - (\der \proj{\adj} V) \cdot 
\end{equation*}
which acts on $E$ and other generalised tensor bundles for the group $G$. We require this to be naturally well-defined (diffeomorphism and gauge invariant in physics terminology) and to satisfy the Leibniz identity
\begin{equation*}
	[\Lgen_V, \Lgen_{V'}] = \Lgen_{\Lgen_V V'} 
\end{equation*}
so that it gives $E$ the structure of a Leibniz algebroid. 
Note that all of the local features of our definition are determined purely by the group $G$, its $\GL(d,\bbR)$ subgroup and the representation for $E$. Globally, it could be twisted by the transformations generated by the Dorfman derivative, in a way specified by additional global data.

In practice, we will see in our examples that the tensor bundle parts of $E$ are differential forms or differential forms weighted by positive powers of the top-form line bundle. Also, the $\GL(d,\bbR)$ decomposition of the adjoint representation of $G$ will feature only $p$-forms and $p$-vectors (as in the statement of~\cite{Baraglia}), so that only the differential form parts of $E$ play an active role in the associated Dorfman derivatives. As we will explain later, this is closely connected to the diffeomorphism covariance of the Dorfman derivative. Overall, this matches well with the generators of the diffeomorphism and gauge symmetries and the field content of standard supergravity theories, and so we will effectively include these forms of the decompositions in our requirements on $E$ for the purposes of the present paper.

Examples of such groups and decompositions were presented in~\cite{Baraglia}, based on the $B$, $D$ and $E$ series of Lie algebras. In this paper we will provide new classes of examples, and explain how they appear in supergravity.

In particular we will present a new series of generalised geometries based on the groups $\Spin(d,d)\times\bbR^+$, with the generalised tangent space transforming as a spinor representation. This will include a $(d-2)$-form potential in the geometry. We we also mention a similar series based on the group $\SL(d+1,\bbR)\times\bbR^+$, which will include a $(d-1)$-form potential. In this case the generalised tangent space will be the antisymmetric bi-vector representation.

The algebras we study here will all correspond to real forms of a Dynkin diagram with a so-called gravity line of nodes associated to a $\GL(d,\bbR)$ subalgebra, as in~\cite{Kleinschmidt:2003mf}. We consider only finite dimensional algebras and always include an overall $\bbR^+$ factor as in~\cite{CSW1,CSW2,CSW3}. We label the standard representation of the $\GL(d,\bbR)$ subalgebra as $T$,\footnote{We slightly abuse notation in not distinguishing carefully between this representation and the tangent bundle of a $d$-dimensional manifold in a way which should not cause confusion.} using the convention that $T$ corresponds to the Dynkin node at the left end of the gravity line, while the node at the right end corresponds to $T^*$. 
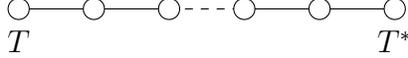
\begin{center}
\begin{tikzpicture} [place/.style={circle,draw=black,fill=white, inner sep=0pt,minimum size=8}] 
\draw (-1,0)--(0,0); 
\draw (0,0)--(1,0);
\draw[dashed] (-2,0)--(-1,0);
\draw (-2,0)--(-3,0);
\draw (-3,0)--(-4,0);
\ph{\node at (0,0.5) [place,label=below:$$] {};}
\node at (0,0) [place,label=below:$$] {}; 
\node at (1,0) [place,label=below:$T^*$] {}; 
\node at (-1,0) [place,label=below:$$] {}; 
\node at (-2,0) [place,label=below:$$] {};
\node at (-4,0) [place,label=below:$T$] {};
\node at (-3,0) [place,label=below:$$] {};
\end{tikzpicture}
\end{center}
This distinction will prove to be important in constructing our new examples of generalised geometries. In a sense, we will simply reverse the orientation of the gravity line of some previously known cases.

To this gravity line can be attached other nodes. For example, one could attach a node with a single line to the $p^{\text{th}}$ node from the right
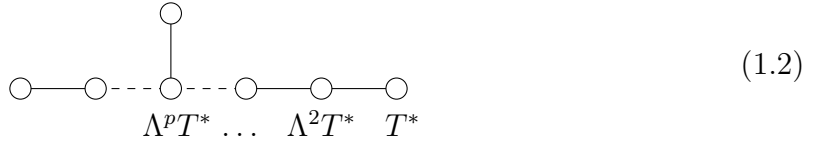
\begin{equation}
\label{eq:p-form-diagram}
\begin{tikzpicture} [place/.style={circle,draw=black,fill=white, inner sep=0pt,minimum size=8},baseline=(current  bounding  box.center)]  
\draw[dashed] (-1,0)--(0,0); \draw(-1,0)--(-1,1);
\draw (0,0)--(1,0);
\draw (1,0) -- (2,0); 
\draw[dashed] (-2,0)--(-1,0);
\draw (-2,0)--(-3,0);
\node at (2,0) [place,label=below:$\ph{{}^2}T^*$] {};
\node at (0,0) [place,label=below:$\ph{I} \dots \ph{I^2}$] {}; 
\node at (1,0) [place,label=below:$\ph{}\Lambda^2 T^*$] {}; 
\node at (-1,0) [place,label=below:$ \ph{{}^2} \Lambda^p T^*$] {}; 
\node at (-1,1) [place,label=right:$$] {}; 
\node at (-2,0) [place,label=below:$$] {};
\node at (-3,0) [place,label=below:$$] {};
\end{tikzpicture}
\end{equation}
Schematically, this will add a generator of the form $\Lambda^p T \oplus \Lambda^p T^*$ to the $\GL(d,\bbR)$ decomposition of the adjoint. 
Such a term in the adjoint representation is related to a $p$-form potential in the corresponding gravitational theory. This pattern holds for zero-form and top-form potentials, for which there is no associated Dynkin node so one adds an $\SL(2,\bbR)$ factor, and also for more exotic fields such as the dual graviton of~\cite{Curtright:1980yk,west-conj,Hull:2001iu}. 
In the simplest cases, the adjoint representation from~\eqref{eq:p-form-diagram} will simply become
\begin{equation}
\label{eq:p-form-adj}
	\adj \ra (T \otimes T^*) \oplus \Lambda^p T \oplus \Lambda^p T^*
\end{equation}
though in general there will be additional generators which arise from commutators of these ones, and these must be analysed in each particular case.  
We devote appendix~\ref{app:examples} to exploring these patterns by means of several examples, with references to the literature as all of these examples have appeared before. 
However, 
the new geometries we will introduce 
in our main discussion 
simply have a diagram of the form~\eqref{eq:p-form-diagram} 
and a decomposition of the adjoint representation~\eqref{eq:p-form-adj}. 
Thus, even the most basic cases can lead to new examples and such patterns are useful for inspiring these constructions.

We observe that the Dynkin label corresponding to the generalised tangent space always has the form $[1, 0, \dots , 0; *]$ where the labels before the semi-colon are those of the gravity line. The embedding of $\GL(d,\bbR)$ in the enlarged algebra is defined so that the decomposition of this representation has the form $T\oplus (\dots)$. We will draw the Dynkin diagrams with the nodes corresponding to the generalised tangent space labelled with an $E$. In fact, the representation theoretical structure of the so-called ``section condition"~\cite{Hohm:2010jy,BP-alg,CSW2,West:2012qm} (or rather the complementary irreducible parts of $S^2 E$) can also be read off from looking at Dynkin labels. This is described in appendix~\ref{app:section}. The $T$ part of the generalised tangent space is stabilised by a parabolic subgroup. Moreover, any subspace which is null in the section condition is also stabilised by such a subgroup. The corresponding parabolic subalgebra was described in~\cite{Baraglia}. The parabolic subalgebras are in one-to-one correspondence with the set of subsets of nodes of the Dynkin diagram, the one of relevance here corresponding to the gravity line. Note that if the gravity line corresponds to a non-maximal $\GL(d,\bbR)$ subalgebra, then the null subspace is also not maximal. This occurs, for example, in the type II decompositions of~\cite{CSW2}.

Most such diagrams that one can draw do not give rise to generalised geometries. This is because of the appearance of tensor fields with mixed Young tableaux symmetry, as in~\cite{Curtright:1980yk}, such as the dual graviton. As the non-linear construction of physical theories based on these types of fields is highly problematic, it is not surprising to find that the simple generalised geometry construction fails in these cases. The central problem here is that the Dorfman derivative fails to be covariant under diffeomorphisms\footnote{
There are also algebraic issues (see e.g.~\cite{BCKT}), which may be cured~\cite{Hohm:2013jma} by an approach inspired by considering the tensor hierarchy~\cite{deWit:2008ta} of the external space theory.
}. This is due to the absence of a diffeomorphism covariant notion of gauge transformations for these mixed symmetry fields. We will deliberately endeavour to avoid these fields throughout this paper, giving only a brief algebraic discussion in appendix~\ref{app:examples}. References on this include~\cite{Hohm:2013jma,West:2002jj,MixedSym1,MixedSym2,MixedSym3,MixedSym4,MixedSym5,MixedSym6,MixedSym7,MixedSym8,MixedSym9,MixedSym10,MixedSym11,MixedSym12}.

However, if the decomposition of the adjoint representation contains only $T\otimes T^*$ (the $\GL(d,\bbR)$ subalgebra) and pairs of the type $\Lambda^p T \oplus \Lambda^p T^*$, then the algebra will give rise to a generalised geometry. This fact was observed in~\cite{Baraglia} and corresponds to the fact that the projection which defines the Dorfman derivative (see~\cite{CSW2}) is diffeomorphism covariant if it only involves the exterior derivative and Lie derivative.

A point that we will pick up on in this paper is the idea of considering geometries built from subalgebras of the full continuous ``U-duality"~\cite{HT} algebra. In particular we choose subalgebras of the type described above, and these will geometrise only a subsector of the field content. Indeed the original $O(d,d)$ generalised geometry~\cite{GCY,Gualtieri,CSW1} includes only the NS-NS sector of the field content of type II supergravity. There are cases where the full algebra does not give rise to a geometry, but the subalgebra does. The $\Spin(8,8)\times\bbR^+$ geometry in section~\ref{sec:spindd} provides an example of this, as it is a subalgebra of $E_{8(8)}\times\bbR^+$, for which there is no corresponding geometry. Another example is sketched in appendix~\ref{app:E77-IIA}.

We conclude this introductory section with a brief discussion of how all of this fits into the literature on hidden symmetries in supergravity. Firstly, we note that the connections between algebras of the types described above and supergravity has a long history. The appearance of such symmetries goes back to~\cite{Julia1,Julia2,Julia3} and was further developed in~\cite{deWN1,deWN2,deWN3,deWN4,deWN5,deWN6,deWN7,deWN8,deWN9,deWN10}. The idea that integral exceptional groups could be exact symmetries of quantised string theory was first proposed in~\cite{HT}.

Later, much grander proposals emerged of how infinite dimensional algebras could underly eleven-dimensional supergravity and M theory~\cite{west-conj,west1,west2,E10-1,E10-2}. A more systematic investigation of their appearance and the identification of the various terms appearing at low levels in the decompositions was performed in~\cite{West:2002jj,Nicolai:2003fw,Kleinschmidt:2003mf} (see also an earlier work~\cite{Cremmer:1999du} which considers the finite dimensional cases). (We emphasise that much of the above schematic discussion of the structure of the algebras is contained in these references as well as far more rigorous details.) This was continued in~\cite{Riccioni:2006az}, where interpretations were found for some of the higher level terms, arguing that infinitely many of them are higher dual versions of the original supergravity fields. Throughout the present work, we will refer to these as ``higher duals" though we will not discuss them beyond their appearance in certain algebraic decompositions. Similar algebraic constructions for type II~\cite{Schnakenburg:2001he,West:2004st}, half-maximal~\cite{Schnakenburg:2004vd} and also eight supercharge theories~\cite{Riccioni:2008jz, Houart:2009ed} have been worked out. 

One purpose of the present paper is to explore generalised geometries based around (the finite dimensional cases of) these algebraic constructions. In particular, we wish to describe the dynamics geometrically using the Dorfman derivative, where the above references consider non-linear realisations.

This paper is organised as follows. In section~\ref{sec:spindd} we introduce $\Spin(d,d)\times\bbR^+$ generalised geometry and its appearance in supergravity. In section~\ref{sec:tA6} we discuss a series of ``half-exceptional" geometries, which correspond to a subsector of the full $\Edd\times\bbR^+$ geometries including only the six-form gauge field. There it is seen how the $\Spin(8,8)\times\bbR^+$ geometry provides the $d=8$ case of this series, and supersymmetry variations are derived from it. In these two main sections, the general prescription for the geometry is exactly that of~\cite{CSW2,CSW3}, to which we refer the reader for an explanation of the overall logic of the construction. For this reason, the discussion is not as explicit as that in~\cite{CSW2,CSW3} and we will merely state the results, but the details are straightforward to derive. Section~\ref{sec:conc} contains some discussion of our findings.

Appendix~\ref{app:A} specifies our conventions and also gives some technical details related to the closure of the algebra of the Dorfman derivative from section~\ref{sec:spindd}. In appendix~\ref{app:section}, we show how to find the ``section condition" for an arbitrary Dynkin diagram. Appendix~\ref{app:examples} contains a survey of decompositions of other algebras, most of which do not give rise to geometries, but which complement the discussion in the main text.


\section{$\Spin(d,d)\times\bbR^+$ generalised geometry}
\label{sec:spindd}


$\Spin(d,d)\times\bbR^+$ generalised geometry is the generalised geometry based upon the diagram
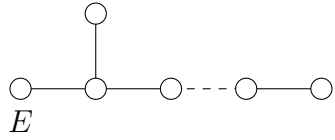
\begin{center}
\begin{tikzpicture} [place/.style={circle,draw=black,fill=white, inner sep=0pt,minimum size=8}] 
\draw[dashed] (-1,0)--(0,0); \draw(-2,0)--(-2,1);
\draw (0,0)--(1,0);
\draw(-2,0)--(-1,0);
\draw (-2,0)--(-3,0);
\node at (1,0) [place,label=below:$$] {};
\node at (0,0) [place,label=below:$$] {}; 
\node at (-1,0) [place,label=below:$$] {}; 
\node at (-2,1) [place,label=right:$$] {}; 
\node at (-2,0) [place,label=below:$$] {};
\node at (-3,0) [place,label=below:$E$] {};
\end{tikzpicture}
\end{center}
As in the introduction, this indicates that the structure group of the geometry is $\Spin(d,d)\times\bbR^+$ and the fibre of the generalised tangent space is the fundamental representation corresponding to the node labelled $E$, which is in this case one of the spinor representations. This is very different to the $O(d,d)$ generalised geometry of~\cite{GCY,Gualtieri}, which would correspond to the diagram
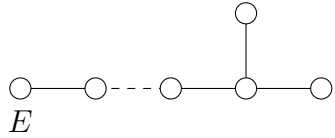
\begin{center}
\begin{tikzpicture} [place/.style={circle,draw=black,fill=white, inner sep=0pt,minimum size=8}] 
\draw(-1,0)--(0,0); \draw(2,0)--(2,1);
\draw[dashed] (0,0)--(1,0);
\draw (2,0) -- (3,0); 
\draw (1,0) -- (2,0);
\node at (1,0) [place,label=below:$$] {};
\node at (2,0) [place,label=below:$$] {}; 
\node at (3,0) [place,label=below:$$] {};
\node at (0,0) [place,label=below:$$] {}; 
\node at (-1,0) [place,label=below:$E$] {}; 
\node at (2,1) [place,label=right:$$] {}; 
\end{tikzpicture}
\end{center}
though, due to $\Spin(4,4)$ triality, the two geometries coincide for $d=4$.

One instance of this geometry has appeared in the literature before, as the case $d=5$ coincides with $E_{5(5)}\times\bbR^+$ generalised geometry~\cite{Hull07,CSW2,BP2} which is relevant to eleven-dimensional supergravity on five-dimensional spaces. Here we will describe these geometries more generally, with particular interest in the case of $d=8$, as this describes ``half" of $E_{8(8)}\times\bbR^+$ in a way which will be described in section~\ref{sec:tA6}.


\subsection{Algebraic decompositions under $\GL(d,\bbR)$}
\label{sec:spindd-decomp}

The first step in the analysis here is to look for the desired embedding of $\GL(d,\bbR)$ which gives a $(d-2)$-form in the decomposition of the adjoint of $\Spin(d,d)$. With the embedding of~\cite{Gualtieri}, one has:
\begin{equation}
	\adj (\Spin(d,d)) \ra (W \otimes W^*) \oplus \Lambda^2 W \oplus \Lambda^2 W^*
\end{equation}
where $W$ is the standard representation of this $\GL(d,\bbR)$ subgroup. Consider setting
\begin{equation}
	 \Lambda^2 W = \Lambda^{(d-2)} T^* = \Lambda^d T^* \otimes \Lambda^2 T
\end{equation}
where $T$ is also a fundamental representation of $\GL(d,\bbR)$ but with a different weight under the $\bbR^+_{\text{diagonal}} \subset \GL(d,\bbR)$. This leads to the identification
\begin{equation}
\label{eq:WvsT}
	 W =  (\Lambda^d T^*)^{\tfrac12} \otimes T
\end{equation}
We then have
\begin{equation}
	\adj (\Spin(d,d)) \ra (T \otimes T^*) \oplus \Lambda^{(d-2)} T \oplus \Lambda^{(d-2)} T^*
\end{equation}
which is the desired decomposition. Henceforth, we will consider the $\GL(d,\bbR)$ subgroup which acts naturally on $T$ to be the one of relevance. This switching of the choice of $\GL(d,\bbR)$ subgroup inside $\Spin(d,d) \times \bbR^+$ is essentially the reversal of the gravity line in the diagram. We note here that the parabolic subalgebra, which will correspond to the geometric subgroup in the context of generalised geometry, is spanned by the subspace 
\begin{equation}
\label{eq:parabolic}
	\adj(\GL(d,\bbR)) \oplus \Lambda^{(d-2)} T^*
\end{equation}
which will correspond to diffeomorphisms and $(d-2)$-form gauge transformations in the physics.

As in~\cite{CSW1,CSW2,CSW3}, the embedding of this $\GL(d,\bbR)$ subgroup will involve a non-trivial $\bbR^+$ factor part in the full structure group $\Spin(d,d)\times\bbR^+$. We make the definition
\begin{equation}
\label{eq:unit-weight}
	\rep{1}_{+1} \simeq (\Lambda^d T^*)^{\tfrac{d-4}{4}} ,
\end{equation}
the appropriateness of which will become apparent when we see that the generalised tangent space will have unit weight under the $\bbR^+$ factor.

We now turn to the decomposition of the spinor representation which will be the fibre of the generalised tangent space. As the chirality of the spinor depends on whether $d$ is odd or even, we treat these cases separately.

For $d$ odd, the spinor has positive chirality. As in~\cite{Gualtieri}, we have the decomposition of the weight zero, positive chirality spinor of $\Spin(d,d)\times\bbR^+$ as
\begin{equation}
\begin{aligned}
	S^+_{0} \ra &(\Lambda^d W)^{\tfrac12} \otimes \Big[ \Lambda^{(\text{even})} W^* \Big]
\end{aligned}
\end{equation}
By~\eqref{eq:WvsT}, this leads to the decomposition of the weight one spinor $S^+_{+1} = S^+_0 \otimes \rep{1}_{+1}$
\begin{equation}
\begin{aligned}
	S^+_{+1} \ra \,
		& T \oplus \Lambda^{(d-3)} T^* \oplus (\Lambda^d T^* \otimes \Lambda^{(d-5)} T^*)
			\oplus ((\Lambda^d T^*)^2 \otimes \Lambda^{(d-7)} T^*) \\ 
			& \hspace{230pt}
			\oplus \dots \oplus ((\Lambda^d T^*)^{(d-3)/2})
\end{aligned}
\end{equation}

Conversely, for $d$ even, the spinor has negative chirality. By similar means, we arrive at the decomposition of the weight one spinor $S^-_{+1} = S^-_0 \otimes \rep{1}_{+1}$
\begin{equation}
\begin{aligned}
	S^-_{+1} \ra \,
		& T \oplus \Lambda^{(d-3)} T^* \oplus (\Lambda^d T^* \otimes \Lambda^{(d-5)} T^*)
			\oplus ((\Lambda^d T^*)^2 \otimes \Lambda^{(d-7)} T^*) \\ 
			& \hspace{220pt}
			\oplus \dots \oplus ((\Lambda^d T^*)^{(d-4)/2} \otimes T^*)
\end{aligned}
\end{equation}
%


\subsection{$\Spin(d,d)\times\bbR^+$ generalised tangent space and generalised tensors}

One can now exactly follow through the construction of~\cite{CSW1,CSW2,CSW3} with these algebras and representations. One now thinks of $T$ as the tangent bundle of a $d$-dimensional manifold and considers a generalised tangent space $E$ as a bundle with a local isomorphism
\begin{equation}
	E \simeq T \oplus \Lambda^{(d-3)} T^* \oplus (\Lambda^d T^* \otimes \Lambda^{(d-5)} T^*)
			\oplus ((\Lambda^d T^*)^2 \otimes \Lambda^{(d-7)} T^*) \oplus \dots
\end{equation}
on patches of the manifold. On the overlaps of patches, one has transition functions given by diffeomorphisms and gauge transformations, the action of the latter being the $\Spin(d,d)\times\bbR^+$ action of exponentiated exact $(d-2)$-forms. The structure group of the generalised tangent bundle is thus the parabolic subroup of $\Spin(d,d)\times\bbR^+$ generated by the subalgebra~\eqref{eq:parabolic}.

However, one can still construct a $\Spin(d,d)\times\bbR^+$ frame bundle, in the same way that an $\Edd\times\bbR^+$ frame bundle was constructed in~\cite{CSW2}, by acting with local $\Spin(d,d)\times\bbR^+$ transformations on the natural local frames induced by coordinates. This is then a $\Spin(d,d)\times\bbR^+$ principal bundle, which enables us to construct $\Spin(d,d)\times\bbR^+$ vector bundles with any representation as the fibre. These are the generalised tensor bundles for the geometry.

A generic $\Spin(d,d)\times\bbR^+$ frame $\{ \hE_\alpha \}$ carries a spinor index $\alpha = 1, \dots , 2^{d-1}$ and we can express a generalised vector as $V = V^\alpha \hE_\alpha$. In even dimensions, $E^*$ has the representation $S^-_{-1}$ as its fibre, so one can write a dual basis with the same spinor index $\{ E^\alpha \}$. In odd dimensions, the fibre of $E^*$ is $S^-_{-1}$, which carries the other spinor index to that for $E$. The dual basis therefore is written as $\{ E^{\dot{\alpha}} \}$, where also $\dot{\alpha} = 1, \dots , 2^{d-1}$.

We will primarily focus on the example of $d=8$ in this paper, so from now on for notational convenience we restrict focus to the case of $d$ even, though of course very similar statements also hold for the case of $d$ odd.


\subsection{The Dorfman derivative and the bundle $N$}

The Dorfman derivative by a generalised vector $V\in E$ can be defined using the definition of~\cite{CSW2}
\begin{equation}
\label{eq:Dorfman-abs}
	\Lgen_V = \der_V - (\der \proj{\adj} V) \cdot ,
\end{equation}
where, as usual in generalised geometry, the partial derivative is promoted to have an $E^*$ index using the embedding $T^* \ra E^*$. As in~\cite{CSW2}, the symbol $\proj{\adj}$ indicates the projection of the partial derivative of the components of $V$, which has the indices of $E^* \otimes E$, onto the adjoint of $\Spin(d,d)\times\bbR^+$.

One can see immediately that the Dorfman derivative will be covariant under diffeomorphisms by examining the $\GL(d,\bbR)$ decompositions of $E$ and $E^*$. Roughly, the Dorfman derivative is a combination of the Lie derivative along the $T$ direction in $E$ and the $\Spin(d,d)$ action of the $(d-2)$-form $\dd \omega$, where $\omega$ is the $\Lambda^{d-3} T^*$ part of $V$. No other contributions to the second term of~\eqref{eq:Dorfman-abs} are compatible with $\GL(d,\bbR)$.

The Dorfman derivative is most usefully written in spinor indices. Acting on another generalised vector $W = W^\alpha \hE_{\alpha}$, we have\footnote{Recall that we are taking $d$ even here.}
\begin{equation}
\label{eq:spinddDorfman}
	(\Lgen_V W)^\alpha = V^\beta \der_\beta W^\alpha 
		+ \tfrac18 (\sigma_{MN})^\gamma{}_\delta (\der_\gamma V^\delta) 
			(\sigma^{MN})^\alpha{}_\beta W^\beta
		+ \tfrac{d-4}{4} (\der_\beta V^\beta) W^\alpha
\end{equation}
where here the matrices $\sigma_{MN}$ are the generators of the $\Spin(d,d)$ algebra acting on the spinors $V^\alpha$. For more details of our conventions, see appendix~\ref{app:conv}. One can also act on other generalised tensors, for example a generalised tensor $X$ transforming in the vector representation of $\Spin(d,d)$ with zero weight under $\bbR^+$. This Dorfman derivative can be written as
\begin{equation}
\label{eq:LVX}
	(\Lgen_V X)^M = V^\alpha \der_\alpha X^M 
		+ \tfrac12 (\sigma^M{}_N)^\alpha{}_\beta (\der_\alpha V^\beta) X^N
\end{equation}

One can then study the closure of the algebra of the Dorfman derivative. To do this using the expressions with spinor indices above, one needs to make note of some combinations of two partial derivatives which vanish identically, due to the fact that only the components of $\der_\alpha$ along $T^*$ are non-vanishing. In fact, studying the $\GL(d,\bbR)$ decompositions, one finds that only the irreducible parts 
\begin{equation}
\label{eq:spinddsec1}
	(\sigma_{M_1 \dots M_{d-2}})^{[\alpha \beta]} \der_\alpha (\dots) \der_\beta (\dots)
	\qquad \text{and} \qquad
	(\sigma_{M_1 \dots M_{d}})^{(\alpha \beta)} \der_\alpha (\dots) \der_\beta (\dots)
\end{equation}
of two separate derivatives and, for second derivatives, only
\begin{equation}
\label{eq:spinddsec2}
	(\sigma_{M_1 \dots M_{d}})^{(\alpha \beta)} \der_\alpha \der_\beta (\dots)
\end{equation}
can be non-vanishing. The remaining irreducible parts of $S^2 E^*$ form the bundle $N^* \subset S^2 E^*$, whose dual $N$ is the $\Spin(d,d)\times\bbR^+$ version of the bundle $N$ from~\cite{CSW2}, which governs the ``section condition" of extended geometries. A general method to identify this bundle can be found in appendix~\ref{app:section}.

Armed with~\eqref{eq:spinddsec1} and~\eqref{eq:spinddsec2}, one can see the closure of the algebra using Fierz identities. Some of the steps of this derivation are highlighted in appendix~\ref{app:Fierz}. In fact, the closure of the algebra is guaranteed by the results of~\cite{Baraglia}, and the structure forms a Leibniz algebroid.


\subsection{Generalised connections and torsion}

Generalised connections are defined simply as linear differential operators
\begin{equation}
	\Dgen : B \ra E^* \otimes B
\end{equation}
where $B$ is any $\Spin(d,d)\times\bbR^+$ tensor bundle and the generalised torsion is defined for $V\in E$ by
\begin{equation}
\label{eq:torsion-abs}
	T(V) = \Lgen^{(\Dgen)}_V - \Lgen_V
\end{equation}
acting on any generalised tensor.

Writing $\Dgen_\alpha = \der_\alpha + \Omega_\alpha$, where $\Omega_\alpha$ is a local Lie algebra valued section of $E^*$, one can see that $\Omega$ has a decomposition into $\Spin(d,d)\times\bbR^+$ irreducible parts
\begin{equation}
	S^-_{-1} \otimes \adj(\Spin(d,d)\times\bbR^+) = S^-_{-1} + S^-_{-1} + K_{-1} + P_{-1}
\end{equation}
where $K$ is the representation corresponding to the positive chirality spin-$\tfrac32$ representation, and $P$ is the remaining irreducible part.

From~\eqref{eq:spinddDorfman}, one can easily see that the generalised torsion lives in the representations $S^-_{-1} \oplus K_{-1}$. The $\GL(d,\bbR)$ decomposition of this contains the terms
\begin{equation}
	S^-_{-1} \oplus K_{-1} \ra T^* \oplus (T \otimes \Lambda^2 T^*) \oplus \Lambda^{d-1} T^* 
		\oplus \dots
\end{equation}
so the generalised torsion contains the ordinary torsion as well as terms for a $(d-1)$-form field strength and the derivative of a scalar.


\subsection{Split frames and $\Spin(d)\times\Spin(d)$ structures}

As in~\cite{CSW1,CSW2}, one can construct so-called conformal split frames for the geometry, essentially by acting on a local coordinate induced frame $\{ \hE_{\alpha} \} = \{ \der / \der x^m, \dd x^m_1 \wedge \dots \wedge \dd x^{m_{d-3}}, \dots \}$ with an element of the geometric subgroup, which untwists the patching of the generalised tangent space, and an $\bbR^+$ scaling. The key ingredient of this group element is a $(d-2)$-form gauge field which has the same gauge transformation patching as the twisting of the generalised tangent space. The split frames concretely realise the global isomorphism
\begin{equation}
	E \simeq T \oplus \Lambda^{(d-3)} T^* \oplus (\Lambda^d T^* \otimes \Lambda^{(d-5)} T^*)
			\oplus ((\Lambda^d T^*)^2 \otimes \Lambda^{(d-7)} T^*) \oplus \dots
\end{equation}

Now suppose we have a metric $g_{mn}$, a scalar field $\Delta$ and a $(d-2)$-form gauge field $A_{m_1 \dots m_{d-2}}$. One can build a particular $\SO(d)$ family of split frames corresponding to these fields, by applying the untwisting transformation (by $A_{(d-1)}$) and $\bbR^+$ scaling (by $\ee^{\Delta}$) to the coordinate induced frame on $E$ as above, and then working in a vielbein frame $\hat{e}_a{}^m$ for the given metric on the tangent bundle $T$.

In one of these split frames $\{ \hE_\alpha \}$, one can define a positive definite inner product on $E$ by 
\begin{equation}
	G(V,V) = \delta_{ab} V^a V^b 
		+ \tfrac{1}{(d-3)!} \delta^{a_1 b_1} \dots \delta^{a_{d-3} b_{d-3}}
			V_{a_1 \dots a_{d-3}} V_{b_1 \dots b_{d-3}}
		+ \dots
\end{equation}
where $V = V^a \hE_a + \tfrac{1}{(d-3)!} V_{a_1 \dots a_{d-3}} \hE^{a_1 \dots a_{d-3}} + \dots$. This inner product is stabilised by $\Spin(d)\times\Spin(d)$, the maximal compact subgroup of $\Spin(d,d)\times\bbR^+$, so the generalised vielbein frames for this generalised metric form a $\Spin(d)\times\Spin(d)$ structure.

Given this structure, one can then go through the remaining steps in the construction of~\cite{CSW1,CSW2,CSW3}. One finds a family of torsion-free compatible connections and a set of unique operators associated to them acting on certain spinor bundles of $\Spin(d)\times\Spin(d)$. Using these, one can look to construct a generalised Ricci curvature tensor as in~\cite{CSW2,CSW3}. We do not give details of this here, but the calculations are straightforward.


\subsection{Appearance in supergravity}

Recall that generalised geometry typically describes the internal sector of compactifications of supergravity. We refer to this internal sector as a dimensional restriction of the original theory. Essentially, one performs a dimensional split taking the external space to be Minkowski, and keeps only fields depending on the internal coordinates which do not break the symmetry of the external space. The exact prescription for dimensional restriction is given in~\cite{CSW3} for the example of eleven-dimensional supergravity restricted to $d$-dimensional spaces, where it is shown to be described by $\Edd\times\bbR^+$ generalised geometry. 

Similarly, for small values of $d$, the $\Spin(d,d)\times\bbR^+$ generalised geometries all appear in maximal supergravity. Here we give a brief discussion of some examples.

\subsubsection*{$d=4$ and $d=5$}

As mentioned before, the $d=4$ case coincides exactly with the original $O(4,4)$ generalised geometry of~\cite{GCY, Gualtieri}, after a $\Spin(4,4)$ triality rotation. One can see that the $\bbR^+$ weight of $E$ vanishes, as does the relevant term of the Dorfman derivative. The relevance of this geometry to type II theories is well-known~\cite{CSW1,gen-susy1,gen-susy2,gen-susy3,gen-susy4,sigmas1,sigmas2,sigmas3,sigmas4,Gen-geom1,Gen-geom2,Gen-geom3,Gen-geom4,Gen-geom5,Gen-geom6,Gen-geom7,Gen-geom8,Gen-geom9,Gen-geom10,Gen-geom11,Gen-geom12,Gen-geom13,Gen-geom14,Gen-geom15,gen-calib1,gen-calib2,gen-calib3,gen-calib4,gen-calib5}.

The $d=5$ geometry is the $E_{5(5)}\times\bbR^+$ generalised geometry of eleven-dimensional supergravity restricted to five-dimensional spaces~\cite{Hull07, BP2, CSW2, CSW3}.

\subsubsection*{$d=6$}

This geometry can be viewed as a subsector of the $E_{7(7)}\times\bbR^+$ generalised geometry of type IIB supergravity restricted to six-dimensional spaces~\cite{Hull07, CSW2, GLSW, E7-flux}. The generalised tangent space has the decomposition
\begin{equation}
	E \simeq T \oplus \Lambda^3 T^* \oplus (\Lambda^6 T^* \otimes T^*)
\end{equation}
thus including the charges of the D3-brane and dual graviton. Note that no gauge transformation associated to the dual graviton is included in the geometry, so that there are no problems with covariance.

\subsubsection*{$d=7$ and $d=8$}

The algebra one would naturally associate to eleven-dimensional supergravity restricted to eight-dimensional manifolds is $E_{8(8)}\times\bbR^+$, whose decomposition under the relevant gravity line subgroup $\GL(8,\bbR)$ will be given in section~\ref{sec:half-E88}. This algebra includes the potential of the dual graviton, so that it does not give rise to a generalised geometry due to the usual problems with covariance of the Dorfman derivative. 
However, the $\Spin(8,8)\times\bbR^+$ subalgebra can be viewed as a truncation of $E_{8(8)}\times\bbR^+$ which keeps only the six-form potential, as will be described in section~\ref{sec:tA6}, and this does give rise to a generalised geometry. 
In this sense, the $d=8$ case geometrises a sector of eleven-dimensional supergravity not previously covered by a geometric construction of this type. The generalised tangent space decomposes as
\begin{equation}
	E \simeq T \oplus \Lambda^5 T^* \oplus (\Lambda^8 T^* \otimes \Lambda^3 T^*) 
		\oplus ((\Lambda^8 T^*)^2 \otimes T^*)
\end{equation}
The additional charges in the geometry are thus the M5-brane, a higher dual of the M5-brane~\cite{Riccioni:2006az} and a higher dual of the graviton (see appendix~\ref{app:dual-gravity}). Again, the gauge transformations associated to the dual charges are not included here.

The $d=7$ case corresponds to half of the IIA circle reduction of the $d=8$ case. Here $E$ has the decomposition
\begin{equation}
	E \simeq T \oplus \Lambda^4 T^* \oplus (\Lambda^7 T^* \otimes \Lambda^2 T^*) 
		\oplus (\Lambda^7 T^*)^2
\end{equation}
so one has the D4-brane, a dual version of the NS5-brane, and also a higher dual of the D0-brane. One can visualise this reduction in Dynkin diagrams by folding up the node at the right end (as in~\ref{app:KK}) and then truncating it.


\section{Half-exceptional generalised geometry}
\label{sec:tA6}


In this section, we show how the $\Spin(8,8)\times\bbR^+$ geometry of the previous section fits into a series of ``half-exceptional" algebras we denote $\Ehfd$, listed in table~\ref{tab:halfexceptional}. These algebras are constructed by taking the level decompositions\footnote{In the extra node added to the gravity line as in~\cite{Nicolai:2003fw}.} of the exceptional algebras and truncating to even levels only. As the grading respects this operation, the resulting algebra is guaranteed to close. The Dynkin diagrams of the resulting series of algebras closely resemble those of the exceptional algebras, in that there is a gravity line with one node added. However, this node is now added above the sixth node from the right instead of the third. The $\Spin(d,d)\times\bbR^+$ series of the previous section was built by adding nodes to the right end of the Dynkin diagram, which changed the relevant higher dimensional theory as well as the dimension of restriction. The present series adds nodes to the left end, which keeps the higher dimensional theory the same, while increasing the dimension of restriction.

\newcolumntype{V}{>{\centering\arraybackslash} m{.4\linewidth} }
\begin{table}[htb]
\begin{center}
\begin{tabular}{m{6pt}| llV}
   $d$ & $\Ehfd\times\bbR^+$ & $\Hhfd$ & Dynkin diagram \\
   \hline
    & & & \\
   6 & $\SL(6,\bbR)\times\SL(2,\bbR)\times\bbR^+$ & $SO(6)\times SO(2)$ 
   	& 
\begin{tikzpicture} [place/.style={circle,draw=black,fill=white, inner sep=0pt,minimum size=8}] 
\draw(-1,0)--(0,0); 
\draw (0,0)--(1,0);
\draw (2,0) -- (3,0); 
\draw (1,0) -- (2,0);
\node at (1,0) [place,label=below:$$] {};
\node at (2,0) [place,label=below:$$] {}; 
\node at (3,0) [place,label=below:$$] {};
\node at (0,0) [place,label=below:$$] {}; 
\node at (-1,0) [place,label=below:$E$] {}; 
\node at (-2,1) [place,label=below:$E$] {}; 
\ph{\node at (-4,0) [place,label=below:$$] {};}
\end{tikzpicture}
 \\
   7 & $\SL(8,\bbR) \times \bbR^+$ & $SO(8)$ 
   	& 
\begin{tikzpicture} [place/.style={circle,draw=black,fill=white, inner sep=0pt,minimum size=8}] 
\draw(-1,0)--(0,0); \draw(-2,0)--(-2,1);
\draw (0,0)--(1,0);
\draw (2,0) -- (3,0); 
\draw (1,0) -- (2,0);
\draw(-2,0)--(-1,0);
\node at (1,0) [place,label=below:$$] {};
\node at (2,0) [place,label=below:$$] {}; 
\node at (3,0) [place,label=below:$$] {};
\node at (0,0) [place,label=below:$$] {}; 
\node at (-1,0) [place,label=below:$$] {}; 
\node at (-2,1) [place,label=right:$$] {}; 
\node at (-2,0) [place,label=left:$E$] {};
\ph{\node at (-4,0) [place,label=below:$$] {};}
\end{tikzpicture}
\\
   8 & $\Spin(8,8)\times\bbR^+$  & $\Spin(8)\times\Spin(8)$ 
      &
\begin{tikzpicture} [place/.style={circle,draw=black,fill=white, inner sep=0pt,minimum size=8}] 
\draw(-1,0)--(0,0); \draw(-2,0)--(-2,1);
\draw (0,0)--(1,0);
\draw (2,0) -- (3,0); 
\draw (1,0) -- (2,0);
\draw(-2,0)--(-1,0);
\draw (-2,0)--(-3,0);
\node at (1,0) [place,label=below:$$] {};
\node at (2,0) [place,label=below:$$] {}; 
\node at (3,0) [place,label=below:$$] {};
\node at (0,0) [place,label=below:$$] {}; 
\node at (-1,0) [place,label=below:$$] {}; 
\node at (-2,1) [place,label=right:$$] {}; 
\node at (-2,0) [place,label=below:$$] {};
\node at (-3,0) [place,label=below:$E$] {};
\ph{\node at (-4,0) [place,label=below:$$] {};}
\end{tikzpicture}
 \\
   9 & $E_{9(9)}$ & $KE_{9(9)}$ &
\begin{tikzpicture} [place/.style={circle,draw=black,fill=white, inner sep=0pt,minimum size=8}] 
\draw(-1,0)--(0,0); \draw(-2,0)--(-2,1);
\draw (0,0)--(1,0);
\draw (2,0) -- (3,0); 
\draw (1,0) -- (2,0);
\draw(-2,0)--(-1,0);
\draw (-2,0)--(-3,0);
\draw(-3,0)--(-4,0);
\node at (1,0) [place,label=below:$$] {};
\node at (2,0) [place,label=below:$$] {}; 
\node at (3,0) [place,label=below:$$] {};
\node at (0,0) [place,label=below:$$] {}; 
\node at (-1,0) [place,label=below:$$] {}; 
\node at (-2,1) [place,label=right:$$] {}; 
\node at (-2,0) [place,label=below:$$] {};
\node at (-3,0) [place,label=below:$$] {};
\node at (-4,0) [place,label=below:$E$] {};
\end{tikzpicture}
\end{tabular}
\end{center}
\caption{Series of half-exceptional groups and their maximal compact subgroups\label{tab:halfexceptional}}
\end{table}

The representation corresponding to the generalised tangent space is given by a Dynkin label with a $1$ for the nodes labelled $E$ and zero for the other nodes. This overall pattern is exactly as for the exceptional groups. Note that though we added $E_{9(9)}$ in the last line to continue the algebraic pattern, we do not discuss this group further in this paper.

For the supergravity, the truncation to even levels means that one restricts to a subsector of the field content. The full exceptional algebra is generated by the $\Lambda^3 T \oplus \Lambda^3 T^*$ part of the algebra and multiple commutators. This roughly corresponds to the presence of the three-form gauge field $A_{(3)}$ in the supergravity. In the same way, the truncation to even levels is generated instead by the $\Lambda^6 T \oplus \Lambda^6 T^*$ part, which corresponds to the six-form $\tA_{(6)}$. Therefore, it makes perfect sense that these algebras geometrise the subsector consisting of the metric, the six-form and, as the dimension increases, their higher rank dual fields (in the sense of~\cite{Riccioni:2006az}).

Much of this section is concerned with repeating the construction of~\cite{CSW2,CSW3} for these half-exceptional geometries. Therefore, we will mostly state the results, referring the reader to~\cite{CSW2,CSW3} for more explanation of the overall logic.


\subsection{Half-exceptional geometry for $d\leq7$}

The complete description of the half-exceptional geometries for $d\leq7$ can almost be read-off from the equations in~\cite{CSW2,CSW3}, simply by setting the truncated terms to zero. For example the generalised tangent space has a local isomorphism
\begin{equation}
	E \simeq T \oplus \Lambda^5 T^* ,
\end{equation}
the Dorfman derivative becomes
\begin{equation}
\label{eq:Lgen}
\begin{aligned} 
   \Lgen_V V' &= \mathcal{L}_v v' 
       + \left( \mathcal{L}_v \sigma' - i_{v'} \dd\sigma \right) ,
\end{aligned}
\end{equation}
where $v\in T$ and $\sigma\in\Lambda^5 T^*$ are the two parts of the generalised vector $V$, and the generalised torsion acts as
\begin{equation}
\label{eq:splitT}
   T(V) = \ee^\Delta \left(- i_v \dd\Delta + v\otimes \dd\Delta
              - i_v \tF + \dd \Delta \wedge \sigma \right) .
\end{equation}
The $N$ bundle decomposes as
\begin{equation}
\label{eq:N}
\begin{aligned}
   N &\simeq 
       \Lambda^4T^*M \oplus (\Lambda^7T^*M\otimes\Lambda^3T^*M). 
\end{aligned}
\end{equation}
so that the corresponding representation is the fundamental representation for the fourth node from the right of the Dynkin diagram. See~\cite{CSW2} for precise details of the meaning of these expressions.

The maximal compact subgroup $\Hd \subset \Edd \times \bbR^+$ becomes now the maximal compact subgroup $\Hhfd$ of $\Ehfd\times \bbR^+$ as listed in table~\ref{tab:halfexceptional}, which is a subgroup of $\Hd$ with algebra
\begin{equation}
	\adj(\Hhfd) \simeq \Lambda^2 T^* \oplus \Lambda^6 T^* ,
\end{equation}
under an $\SO(d)$ decomposition. The representations of $\Hd$ in which the fermions transform then decompose under $\Hhfd$, but in calculations it is often more convenient to continue to work with the (now reducible) $\Hd$ objects. Note however that there is one substantial simplification in truncating away the $\Lambda^3 T^*$ component of the $\Hd$ algebra: we no longer need to consider the two different representations $S^\pm$ of the algebra on spinors (see~\cite{CSW3}), which were distinguished by the sign of the action of $\Lambda^3 T^*$.

The only equation which must be changed is the expression for the torsion-free compatible connection in the split frame. Recall that in generalised geometry, the torsion-free and compatibility conditions are insufficient to fix the connection uniquely: there are undetermined components. The expression given in~\cite{CSW2} is a particular choice, as the result is ambiguous due to the fact that one could choose to absorb some combinations of terms into the undetermined parts of the connection. This particular choice is no longer available to us in the more restricted setup considered here, but now a valid choice is
\begin{equation}
\begin{aligned}
   \Dgen_a 
      &= \ee^\Delta \left( \nabla_a
          + \tfrac{1}{4} {\left( \tfrac{17-2d}{d-1} \right)} (\der_b \Delta) \gamma_a{}^b
          - \tfrac12 \tfrac{1}{7!} \tF_{ab_1\dots b_6}\gamma^{b_1\dots b_6}
          + \slashed{\am}_a \right), \\
   \Dgen^{a_1\dots a_5} 
      &= \ee^\Delta \left( \tfrac{1}{4} \tfrac{5!}{7!} \tF^{a_1\dots a_5}{}_{b_1b_2}\gamma^{b_1b_2}
      	- \tfrac34 {\tbinom{d-1}{5}}^{-1}
		(\der_b \Delta) \gamma^{ba_1 \dots a_5}
         	+ \slashed{\am}^{a_1\dots a_5}\right) , \\
\end{aligned}
\end{equation}
where again $Q$ represents the parts of the connection which are not determined uniquely.

The unique derivative operators which led to the supersymmetry variations of the fermions in~\cite{CSW3} can be truncated straightforwardly. We reproduce here the relevant terms acting on a spinor $\hat\varepsilon = \ee^{-\Delta/2}  \varepsilon^{\text{sugra}}$, which is promoted to a representation of $\Hhfd$ 
\begin{equation}
\begin{aligned}
   \DSS \hat{\varepsilon}
	&= \Gamma^a \Dgen_a \hat{\varepsilon}
		+ \tfrac{1}{5!} \Gamma^{c_1 \dots c_5} \Dgen_{c_1 \dots c_5} \hat{\varepsilon} \\
      &= \ee^{\Delta/2} \Big( \slashed{\LC} 
         + \tfrac{9-d}{2} (\slashed{\der} \Delta) 
         - \tfrac{1}{4}  \slashed{\tF}  
         \Big) \varepsilon^{\text{sugra}}, \\
   (\DSJ \hat{\varepsilon})_a 
	& = \Dgen_a \hat{\varepsilon}
		- \tfrac13 \tfrac{1}{4!} \Gamma^{c_1 \dots c_4} \Dgen_{ac_1 \dots c_4} \hat{\varepsilon}
		+ \tfrac23 \tfrac{1}{5!} \Gamma_a{}^{c_1 \dots c_5} \Dgen_{c_1 \dots c_5} 
		\hat{\varepsilon} \\
      &= \ee^{\Delta/2} \Big( \LC_a 
      - \tfrac{1}{12} \tfrac{1}{6!} \tF_{ab_1 \dots b_6} 
	 \Gamma^{b_1 \dots b_6} \varepsilon   
         \Big)\varepsilon^{\text{sugra}} .
\end{aligned}
\end{equation}

We briefly note that the $d=7$ case here is part of a family of generalised geometries based on the groups $\SL(d+1,\bbR)\times\bbR^+$, with diagrams
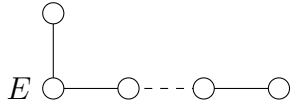
\begin{center}
\begin{tikzpicture} [place/.style={circle,draw=black,fill=white, inner sep=0pt,minimum size=8}] 
\draw[dashed](-1,0)--(0,0); \draw(-2,0)--(-2,1);
\draw (0,0)--(1,0);
\draw(-2,0)--(-1,0);
\node at (1,0) [place,label=below:$$] {};
\node at (0,0) [place,label=below:$$] {}; 
\node at (-1,0) [place,label=below:$$] {}; 
\node at (-2,1) [place,label=right:$$] {}; 
\node at (-2,0) [place,label=left:$E$] {};
\ph{\node at (-4,0) [place,label=below:$$] {};}
\end{tikzpicture}
\end{center}
This family is similar to that of section~\ref{sec:spindd}, but it geometrises a $(d-1)$-form potential, leading to a top-form field strength. Another example of this series is the well-known $E_{4(4)}\times\bbR^+$ generalised geometry in four dimensions studied in~\cite{Hull07,CSW2,CSW3,BP1}. They can be thought of as the ``gravity-line-reversal" of the geometry in appendix~\ref{app:KK}.


\subsection{Half-exceptional geometry for $d=8$: $\Spin(8,8)\times\bbR^+$}
\label{sec:half-E88}

The $\GL(8,\bbR)$ decompositions of the relevant representations of  $E_{8(8)}\times\bbR^+$ read
\begin{equation}
\begin{aligned}
	\rep{1}_{+1} &\ra (\Lambda^8 T^*) \\
	\rep{248}_{0} &\ra (T \otimes T^*) \oplus \Lambda^{3} T \oplus \Lambda^{3} T^*
		\oplus \Lambda^{6} T \oplus \Lambda^{6} T^*
		\oplus (\Lambda^8 T \otimes T) \oplus (\Lambda^8 T^* \otimes T^*) \\
	\rep{248}_{+1} &\simeq \rep{248}_{0} \otimes \rep{1}_{+1} \\
		&\ra T \oplus \Lambda^2 T^* \oplus \Lambda^5 T^* 
			\oplus (T^* \otimes \Lambda^7 T^*) \\
			& \qquad \oplus (\Lambda^8 T^* \otimes \Lambda^3 T^*) 
			\oplus (\Lambda^8 T^* \otimes \Lambda^6 T^*) 
			\oplus ((\Lambda^8 T^*)^2 \otimes T^*) \\
\end{aligned}
\end{equation}
Performing the truncation to even levels on $E_{8(8)}\times\bbR^+$, one is left with the $\Spin(8,8)\times\bbR^+$ subgroup. The geometry we need is thus the $\Spin(8,8)\times\bbR^+$ geometry of the previous section.

The decompositions listed in section~\ref{sec:spindd} provide us with the generalised tangent space and the adjoint bundle associated to the frame bundle. We now look at the decompositions of the bundle $N$ and the torsion representation $K_{-1}$. The fibre of $N$ is the representation $\rep{1}_{+2} \oplus \rep{1820}_{+2}$, so that
\begin{equation}
\begin{aligned}
	N \simeq  (\Lambda^8 T^*)^2 
		\oplus \Lambda^4 T^*  &\oplus (\Lambda^7 T^* \otimes  \Lambda^3 T^* )
		\oplus (\Lambda^8 T^* \otimes \Lambda^2 T^* \otimes \Lambda^6 T^*) \\
		&\oplus ((\Lambda^8 T^*)^2 \otimes T^* \otimes \Lambda^5 T^*)
		\oplus ((\Lambda^8 T^*)^3 \otimes \Lambda^4 T^* )
\end{aligned}
\end{equation}
Note that in order for an expression of the form $\Lgen_V W + \Lgen_W V = \der \proj{E} (V\proj{N} W)$ to exist, one would need a coordinate independent map
\begin{equation}
	\der : (\Lambda^8 T^*)^2 \ra (\Lambda^8 T^*)^2\otimes T^*
\end{equation}
which clearly cannot be canonically defined. Therefore, as for the $E_{7(7)}\times\bbR^+$ geometry of~\cite{CSW2}, no such expression can be written.

The fibre of $K_{-1}$ is the spin-$\tfrac32$ representation $\rep{1920}^+_{-1}$, giving a decomposition
\begin{equation}
\begin{aligned}
	K
	&\simeq (T \otimes \Lambda^2T^*) \oplus \Lambda^7 T^* \\
		&\qquad \oplus (T \otimes \Lambda^4 T)^0 
		\oplus (T^* \otimes \Lambda^6 T)
		\oplus (\Lambda^7 T \otimes  \Lambda^4 T ) \\
		&\qquad \oplus (\Lambda^8 T \otimes T \otimes \Lambda^2 T)^0
		\oplus (\Lambda^8 T \otimes \Lambda^2 T \otimes \Lambda^7 T)
		\oplus ((\Lambda^8 T)^2 \otimes \Lambda^7 T )
\end{aligned}
\end{equation}
At first glance, some of the terms here that would survive on truncating to $d=7$ appear to disagree with those given in~\cite{CSW2}. However, on using the seven-dimensional isomorphism $T^* \otimes \Lambda^6 T = \Lambda^7 T \otimes T^* \otimes T^*=  \Lambda^5 T \oplus (\Lambda^7 T \otimes S^2 T^*)$, one can see that there is no contradiction.

The expressions for the torsion-free compatible connection and unique projections for the supersymmetry variations for $d\leq7$ extend to the case $d=8$ without the need for significant modification. Here there are more parts of the connection to deal with, but we can choose to express the connection (acting on a spinor $\hat\varepsilon = \ee^{-\Delta/2}  \varepsilon^{\text{sugra}}$) as
\begin{equation}
\begin{aligned}
   \Dgen_a 
      &= \ee^\Delta \left( \nabla_a
          + \tfrac{1}{4} {\left( \tfrac{17-2d}{d-1} \right)} (\der_b \Delta) \gamma_a{}^b
          - \tfrac12 \tfrac{1}{7!} \tF_{ab_1\dots b_6}\gamma^{b_1\dots b_6}
          + \slashed{\am}_a \right), \\
   \Dgen^{a_1\dots a_5} 
      &= \ee^\Delta \left( \tfrac{1}{4} \tfrac{5!}{7!} \tF^{a_1\dots a_5}{}_{b_1b_2}\gamma^{b_1b_2}
      	- \tfrac34 {\tbinom{d-1}{5}}^{-1}
		(\der_b \Delta) \gamma^{ba_1 \dots a_5}
         	+ \slashed{\am}^{a_1\dots a_5}\right) , \\
   \Dgen^{(\dots)} 
      &= \ee^\Delta \left( \slashed{\am}^{(\dots)} \right) , \hspace{20pt} \text{for other parts}
\end{aligned}
\end{equation}
so that those terms do not affect the calculation of the projections
\begin{equation}
\label{eq:spin88SUSY}
\begin{aligned}
   \DSS \hat{\varepsilon}
	&= \Gamma^a \Dgen_a \hat{\varepsilon}
		+ \tfrac{1}{5!} \Gamma^{c_1 \dots c_5} \Dgen_{c_1 \dots c_5} \hat{\varepsilon} + (\dots) \\
      &= \ee^{\Delta/2} \Big( \slashed{\LC} 
         + \tfrac{9-d}{2} (\slashed{\der} \Delta) 
         - \tfrac{1}{4}  \slashed{\tF}  
         \Big) \varepsilon^{\text{sugra}}, \\
   (\DSJ \hat{\varepsilon})_a 
	& = \Dgen_a \hat{\varepsilon}
		- \tfrac13 \tfrac{1}{4!} \Gamma^{c_1 \dots c_4} \Dgen_{ac_1 \dots c_4} \hat{\varepsilon}
		+ \tfrac23 \tfrac{1}{5!} \Gamma_a{}^{c_1 \dots c_5} \Dgen_{c_1 \dots c_5} 
		\hat{\varepsilon} + (\dots) \\
      &= \ee^{\Delta/2} \Big( \LC_a 
         +\tfrac16 \tfrac1{7!} \tF_{b_1 \dots b_7} \Gamma_a{}^{b_1 \dots b_7} \varepsilon
      - \tfrac{1}{12} \tfrac{1}{6!} \tF_{ab_1 \dots b_6} 
	 \Gamma^{b_1 \dots b_6} \varepsilon   
         \Big)\varepsilon^{\text{sugra}} ,
\end{aligned}
\end{equation}
as the undetermined pieces of the connection $Q$ cancel.


\subsection{Supersymmetry variations with only $\tF_{(7)}$}

The supersymmetry variation of the eleven-dimensional gravitino can be written in terms of the dual field strength $* \mathcal{F} = * \dd \mathcal{A}_{(3)}$ as
\begin{equation}
	\delta \psi_M = \LC_M \varepsilon + \tfrac{1}{12} \Big[ 
		\tfrac2{7!} (* \mathcal{F})_{N_1 \dots N_7} \Gamma_M{}^{N_1 \dots N_7} 
		- \tfrac1{6!} (* \mathcal{F})_{MN_1 \dots N_6} \Gamma^{N_1 \dots N_6} \Big] \varepsilon
\end{equation}
Using the ansatz
\begin{equation}
	\tF_{m_1 \dots m_7} = * \mathcal{F}_{m_1 \dots m_7}
	\hspace{30pt}
	* \mathcal{F}_{\mu M_1 \dots M_6} = 0
\end{equation}
but otherwise keeping the same reduction of fields as in~\cite{CSW3}, this gives rise to the supersymmetry variations
\begin{equation}
\begin{split}
   \delta \rho &= \left[ 
      \slashed{\LC}
      - \tfrac{1}{4} \slashed{\tilde{F}} 
      + \tfrac{9-d}{2} (\slashed{\der} \Delta) 
      \right] \varepsilon ,\\
   \delta \psi_m &= \left[ 
      \LC_m
      +\tfrac16 \tfrac1{7!} \tF_{n_1 \dots n_7} \Gamma_m{}^{n_1 \dots n_7}
      - \tfrac{1}{12} \tfrac{1}{6!} \tF_{mn_1 \dots n_6} 
	 \Gamma^{n_1 \dots n_6}
      \right] \varepsilon,
\end{split}
\end{equation}
for the fermions in the $d$-dimensional restriction.

These are precisely the expressions reproduced by the projection operators~\eqref{eq:spin88SUSY} in the half-exceptional geometry. The generalised geometry description of supersymmetric backgrounds~\cite{gen-susy1,gen-susy2,gen-susy3,gen-susy4,GLSW,GO1,GO2,GT,CSW4} can therefore be extended to the case of compactifications to three dimensions with only internal $* \mathcal{F}$ fluxes by the $\Spin(8,8)\times\bbR^+$ geometry. Some work examining such backgrounds (as well as more general cases) was presented in~\cite{MicuTalk}.

From this point, one anticipates that the rest of the construction will go through, exactly as in~\cite{CSW2,CSW3}, to provide all of the equations of this restricted theory.


\section{Discussion}
\label{sec:conc}

In this paper, we have constructed a new family of generalised geometries based on the groups $\Spin(d,d)\times\bbR^+$ in which the generalised tangent space corresponds to a spinor representation of the group. We have shown how these geometries arise in supergravity and how the case of $\Spin(8,8)\times\bbR^+$ provides a geometry for a class of supersymmetric backgrounds which fall outside the classes covered previously.

The idea of studying geometries containing subsectors of the field content of a theory is not new, as the original generalised geometry of~\cite{GCY,Gualtieri} covered only the NS-NS sector of type II supergravity. This can be viewed as taking an $O(10,10)\times\bbR^+$ subgroup of $E_{11}$~\cite{West:2010ev}. In a sense, the construction of~\cite{CSW2,CSW3} also contains only a subsector as there the fields are dimensionally restricted. Recently the main focus has been to try to include all of the fields, in increasing dimensions. However, one quickly runs into serious problems even for $E_{8(8)}\times\bbR^+$, related to the problem of dual gravity, and worse still for the infinite-dimensional algebras conjectured to underlie the cases where yet more dimensions  are included, as there are then infinitely many mixed symmetry tensor fields to account for.

While understanding this is obviously an important ultimate goal, it may be worthwhile to study subsectors where the problems associated to these more complicated types of fields do not appear. It seems likely that the $\Spin(d,d)\times\bbR^+$ series will continue to have some role as one includes more dimensions of the eleven-dimensional theory in the geometry. For the case of $\Spin(8,8)$ we have found that the geometric prescription appears to hold good if one simply truncates away the problematic fields. The generalised tangent space still contains the higher level charges, though they do not actively play a role. It seems likely that this pattern will continue. The $\Spin(9,9)$ case contains a six-form charge, which may well be the D6 brane of type IIA restricted to 8 dimensions. More interesting could be the $\Spin(10,10)$ case with a seven-form charge, which could be related to one of the seven-branes in type IIB of~\cite{7Branes1,7Branes2}. The $\Spin(11,11)$ case has an eight-form charge, which may be the totally anti-symmetric part of the dual graviton in the full eleven-dimensions. These cases all deserve some investigation in the future.

The other respect in which it may be useful to consider subsectors is for the study of supersymmetric backgrounds. Clearly, one need not always have all fluxes switched on, so for the purposes of considering backgrounds with only certain fluxes, the analysis could be greatly simplified if one includes only the relevant fluxes in the geometry.

The investigations of appendix~\ref{app:6D} indicate that dimensional restrictions of six-dimensional minimal supergravity can also be described by generalised geometry. Further, one can include vector and tensor mulitiplets in six dimensions, provided the restricted fields parameterise a coset. One encounters the same problems as for $E_{8(8)}\times\bbR^+$ if one tries to include three dimensions or more, but the restrictions to two dimensions appear to work as for $\Edd\times\bbR^+$ for $d\leq7$ in eleven-dimensional supergravity. One can similarly consider $G_{2(2)}\times\bbR^+$ for five-dimensional minimal supergravity restricted to two dimensions and find a similar situation to the $\SO(4,3)\times\bbR^+$ case of~\ref{app:6D}. This suggests that the construction applies to any supergravity theory, so long as the restricted fields parameterise a coset and mixed symmetry tensor fields are not included.

Another overriding question, which we do not attempt to answer here, is what feature of these physical theories causes the appearance of generalised geometry? One could suspect the supersymmetry in supergravity may have a role here, as it seems to be very interwoven in the construction. However, generalised geometry also seems to be applicable in cases with no supersymmetry, and in the case of subsectors it is not clear that the fields considered form supermultiplets, so one can question whether one really has supersymmetry in those cases. Gravity may actually be the only absolutely common ingredient. The answer to this question will hopefully become clearer as more is known about these structures.


\acknowledgments

We would like to thank Daniel Waldram for helpful discussions. 
This work was supported by the German Science Foundation (DFG) under the
Collaborative Research Center (SFB) 676 ``Particles, Strings and the Early Universe''.


\appendix



\section{Conventions and technical details}
\label{app:A}


\subsection{Conventions}
\label{app:conv}

All convention choices whose relevance overlaps with those made in~\cite{CSW2,CSW3} are chosen to match~\cite{CSW2,CSW3}.

We use indices $M,N,\dots = 1, \dots, 2d$ as the vector indices of $\Spin(d,d)$ and spinor indices $\alpha,\beta, \dots = 1, \dots , 2^d$. The generators $\omega_{MN}$ of $\Spin(d,d)$ are taken to acts on vectors and spinors of $\Spin(d,d)$ by
\begin{equation}
	\delta X^M = \omega^M{}_N X^N
	\hspace{30pt}
	\delta V^\alpha = \tfrac14 \omega_{MN} (\sigma^{MN})^\alpha{}_\beta V^\beta
\end{equation}
Where we have spinor inner products given by a real matrix $C_{\alpha \beta}$ below, we use the index conventions
\begin{equation}
	C^{\alpha \beta} = (C^{-1})^{\alpha \beta}
	\hspace{30pt}
	V^\alpha = C^{\alpha \beta} V_\beta
	\hspace{30pt}
	V_\alpha = C_{\alpha \beta} V^\beta
\end{equation}
The contraction of a $\Spin(d,d)\times\bbR^+$ generalised vector $V = V^\alpha \hE_\alpha$ with a generalised dual vector $Z = Z_\alpha E^\alpha$ is defined as $V^\alpha Z_\alpha$. The embeddings are normalised such that if $V$ and $W$ have only vector and one-form parts respectively, then $V^\alpha Z_\alpha = V^m Z_m$.


\subsection{Closure of $\Spin(d,d)\times\bbR^+$ Dorfman algebra and Fierz identities}
\label{app:Fierz}

We examine the algebra of two Dorfman derivatives by $U,V\in E$ acting on $X$ as in~\eqref{eq:LVX}. The interesting point is to see how projections of the partial derivatives have to vanish in order for the terms like $V (\der U) (\der X)$ and $VX (\der \der U)$ to cancel. The former types of terms appear in $([\Lgen_U, \Lgen_V] X - \Lgen_{[U,V]} X)_M$ as
\begin{equation}
\begin{aligned}
	\Big( -\tfrac12 \Big[ \tfrac18 (\sigma^{PQ})^\alpha{}_\beta (\sigma_{PQ})^\gamma{}_\delta
		+ \delta^\alpha{}_\delta \delta^\gamma{}_\beta
		+ \tfrac{d-4}{4} \delta^\alpha{}_\beta \delta^\gamma{}_\delta \Big] 
			V^\beta (\der_\gamma U^\delta) (\der_\alpha X_M) \Big)
				- \Big( U \lra V \Big)
\end{aligned}
\end{equation}
while the latter appear as
\begin{equation}
\begin{aligned}
	\Big( -\tfrac14 (\sigma_{MN})^\alpha{}_\beta \Big[ 
		\tfrac18 (\sigma^{PQ})^\beta{}_\epsilon (\sigma_{PQ})^\gamma{}_\delta
		+ \delta^\beta{}_\delta \delta^\gamma{}_\epsilon
		+ \tfrac{d-4}{4} \delta^\beta{}_\epsilon \delta^\gamma{}_\delta \Big] 
			 (\der_\alpha \der_\gamma U^\delta) V^\epsilon X^N \Big)
				- \Big( U \lra V \Big)
\end{aligned}
\end{equation}
One then applies the Fierz identities detailed below, which show a clear pattern. For the resulting expressions to vanish, one needs that the expressions
\begin{equation}
	(\sigma^{M_1 \dots M_p})^{\alpha\beta} \der_\alpha (\dots) \der_\beta (\dots)
	\qquad \text{and} \qquad
	(\sigma^{M_1 \dots M_q})^{\alpha\beta} \der_\alpha \der_\beta (\dots)
\end{equation}
are non-vanishing only for $p = d-2$ or $p = d$ and $q = d$ respectively. One can see that this is indeed the case by evaluating the decompositions of the relevant representations of $\Spin(d,d)\times\bbR^+$ under $\GL(d, \bbR)$. This also follows directly from the general argument of appendix~\ref{app:section}.

\subsubsection*{Fierz identity for $\Spin(4,4)$}

Here we have a symmetric spinor inner product $C_{(\alpha \beta)}$ on 8-component spinors ($\delta^\alpha{}_\alpha = 8$). We find
\begin{equation}
\begin{aligned}
	\tfrac18 (\sigma^{MN})^\alpha{}_\beta (\sigma_{MN})^\gamma{}_\delta
		+ \delta^\alpha{}_\delta \delta^\gamma{}_\beta
		+ \tfrac{d-4}{4} \delta^\alpha{}_\beta \delta^\gamma{}_\delta 
	= \tfrac18 \Big[ 8 C^{\alpha \gamma} C_{\beta \delta} \Big]
\end{aligned}
\end{equation}

\subsubsection*{Fierz identity for $\Spin(6,6)$}

Here the inner product $C_{[\alpha \beta]}$ is antisymmetric on 32-component spinors and we have
\begin{equation}
\begin{aligned}
	\tfrac18 (\sigma^{MN})^\alpha{}_\beta (\sigma_{MN})^\gamma{}_\delta
		+ \delta^\alpha{}_\delta \delta^\gamma{}_\beta
		+ \tfrac{d-4}{4} \delta^\alpha{}_\beta \delta^\gamma{}_\delta 
	= \tfrac{1}{32} \Big[ - 16 C^{\alpha \gamma} C_{\beta \delta}
		+ \tfrac{8}{2!} (\sigma^{MN})^{\alpha\gamma} (\sigma_{MN})_{\beta\delta} \Big]
\end{aligned}
\end{equation}

\subsubsection*{Fierz identity for $\Spin(8,8)$}

$C_{(\alpha \beta)}$ is symmetric again and we have 128-component spinors. We obtain
\begin{equation}
\begin{aligned}
	&\tfrac18 (\sigma^{MN})^\alpha{}_\beta (\sigma_{MN})^\gamma{}_\delta
		+ \delta^\alpha{}_\delta \delta^\gamma{}_\beta
		+ \tfrac{d-4}{4} \delta^\alpha{}_\beta \delta^\gamma{}_\delta \\
	&\qquad = \tfrac{1}{128} \Big[ 32 C^{\alpha \gamma} C_{\beta \delta}
		+ \tfrac{16}{2!} (\sigma^{MN})^{\alpha\gamma} (\sigma_{MN})_{\beta\delta} 
		+ \tfrac{8}{4!} (\sigma^{M_1 \dots M_4})^{\alpha\gamma} 
			(\sigma_{M_1 \dots M_4})_{\beta\delta}\Big]
\end{aligned}
\end{equation}
%


\section{Section conditions from Dynkin labels}
\label{app:section}

As in the introduction, we order the Dynkin labels so that the first $(d-1)$ places represent the gravity line, while the others correspond to the added nodes, separating the two groups with a semi-colon. The generalised tangent space $E$ then always has a label $[1,0,\dots,0; *]$.

We can then examine the decomposition of the tensor product of two such representations. We find there is always a term with a label $[2,0, \dots 0;*]$ in the decomposition of the symmetric part $S^2 E$, while there is always one of the type $[0,1,0,\dots, 0; *]$ in the antisymmetric part $\Lambda^2 E$. By considering the $\bbR^+_{\text{diagonal}}\subset \GL(d,\bbR)$ weights of the terms in the $\GL(d,\bbR)$ decomposition of $E$, we see that there can only be one term like $S^2 T$ in $S^2 E$ and only one term like $\Lambda^2 T$ in $\Lambda^2 E$. These are always found in the decompositions of the representations with the labels just highlighted. Therefore, if we have two generalised vectors $V$ and $W$ living only in the $T$ part of $E$, then, since $T \otimes T = S^2 T \oplus \Lambda^2 T$, only these irreducible parts of $V \otimes W$ can be non-zero. The bundle labelled $N$ in~\cite{CSW2} therefore corresponds to the sum of
all of the irreducible parts of $S^2 E$ except for the representation with label $[2,0, \dots 0;*]$ identified here.

Considering instead $E^* \simeq T^* \oplus \dots$, one can consider the implications of the above for the partial derivative, which lives only in the $T^*$ component. This allows one to quickly read off which combinations of two partial derivatives must vanish identically. There are two cases of interest.

If both derivatives act on the same object, clearly the antisymmetrised part will vanish. Of the symmetric part, only the irreducible component corresponding to the dual of the $[2,0,\dots, 0;*]$ representation defined above can survive.

If the derivatives act on different objects, then the same component of the symmetric part will survive as for the previous case. However, also only one irreducible component of the antisymmetric part can be non-vanishing: that corresponding to the dual of the $[0,1,0,\dots, 0; *]$ representation defined above. Note that the presence of antisymmetrised derivatives acting on different objects vanishing identically has not been discussed prominently in the literature. This is because in most cases examined so far, the antisymmetric tensor product $\Lambda^2 E$ has been irreducible.

One can also see in the examples of appendix~\ref{app:examples} that the leading $\GL(d,\bbR)$ irreducible components of the bundle $N$ have Dynkin labels which match the gravity line part of the Dynkin label for the containing representation of the enlarged algebra. For example, for the $d\leq7$ geometries in~\cite{CSW2}, the leading component is always $T^*$, while for the $d\leq7$ half-exceptional cases of section~\ref{sec:tA6} it is always $\Lambda^4 T^*$. These are also fairly easy to guess, given the form of $E$.

These mneumonics provide an easy way to find the representation for the bundle $N$, or rather its compliment in $S^2 E$. It seems likely that there is a similar extension of them to find the entire sequence of representations discussed in~\cite{BCKT}, which are related to the tensor hierarchy~\cite{deWit:2008ta}.


\section{Examples of algebras and $\GL(d,\bbR)$ decompositions}
\label{app:examples}


In this appendix we review the gravity line decompositions of some algebras relevant to restrictions of various gravitational theories. We include this to illustrate the general schematic patterns outlined in the discussion in the main text. We will endeavour to point out the references to the literature along the way, and it should be understood that the relevance of these algebras to the physical theories is not new. A important reference for much of the section is~\cite{Kleinschmidt:2003mf}.


\subsection{$\SL(d+1,\bbR) \times \bbR^+$ and Kaluza-Klein reduction}
\label{app:KK}

The most trivial example of an algebra leading to a generalised geometry is $\SL(d+1,\bbR) \times \bbR^+$ with diagram
\begin{center}
\begin{tikzpicture} [place/.style={circle,draw=black,fill=white, inner sep=0pt,minimum size=8}] 
\draw (-1,0)--(0,0); 
\draw (0,0)--(0,1);
\draw[dashed] (-2,0)--(-1,0);
\draw (-2,0)--(-3,0);
\node at (0,1) [place,label=below:$$] {};
\node at (0,0) [place,label=below:$$] {}; 
\node at (-1,0) [place,label=below:$$] {}; 
\node at (-2,0) [place,label=below:$$] {};
\node at (-3,0) [place,label=below:$E$] {};
\end{tikzpicture}
\end{center}
One finds that 
\begin{equation}
	E \simeq T \oplus \bbR
\end{equation}
and
\begin{equation}
	\adj(\SL(d+1,\bbR) \times \bbR^+) \simeq \bbR \oplus (T\otimes T^*) \oplus T \oplus T^*
\end{equation}
This corresponds to $(d+1)$-dimensional gravity restricted to $d$ dimensions. The geometry includes the $d$-dimensional gravity, 1-form gauge field and a scalar. It is clearly very reminiscent of ordinary Kaluza-Klein reduction. When written in $\SL(d+1,\bbR)$ indices, the form of the Dorfman derivative coincides with the ordinary Lie derivative.

The diagram above represents that of ordinary gravity, but with the right-most node ``folded-up" off the gravity line. The pattern can be used fairly generally to examine the $S^1$ reduction of the parent higher dimensional theory. We will see it again below.

A further comment is that the $\SL(d+1,\bbR) \times \bbR^+$ geometry described in section~\ref{sec:tA6} is essentially the ``gravity-line-reversal" of this one.


\subsection{$\SL(d+1,\bbR) \times \bbR^+$ and dual gravity}
\label{app:dual-gravity}

We examine pure $D$ dimensional gravity restricted to $d=D-3$ dimensions (with a warp factor in the metric ansatz). This structure will appear as a subsector in almost all of the rest of the algebras considered in this appendix, so it is natural to study this first.

The relevant algebra is the algebra of the Ehlers group $\SL(d+1,\bbR) \times \bbR^+$. We draw the Dynkin diagram as 
\begin{equation}
\label{eq:dual-grav-diagram}
\begin{tikzpicture} [place/.style={circle,draw=black,fill=white, inner sep=0pt,minimum size=8},baseline=(current  bounding  box.center)] 
\draw(-1,0)--(0,0); \draw(3,0)--(-3,1.3);
\draw (0,0)--(1,0);
\draw (2,0) -- (3,0); 
\draw[dashed] (1,0) -- (2,0);
\draw(-2,0)--(-1,0);
\node at (1,0) [place,label=below:$$] {};
\node at (2,0) [place,label=below:$(d-2)$] {}; 
\node at (3,0) [place,label=right:$(d-1)$] {};
\node at (0,0) [place,label=below:$3$] {}; 
\node at (-1,0) [place,label=below:$2$] {}; 
\node at (-3,1.3) [place,label=left:$E$,label=above:$d$] {}; 
\node at (-2,0) [place,label=left:$E$,label=below:$1$] {};
\ph{\node at (-4,0) [place,label=below:$$] {};}
\end{tikzpicture}
\end{equation}
the numbers indicating the order of the Dynkin labels, which we write as $[n_1, \dots , n_{d-1}; n_d]$.
This setup can be understood as one of a series similar to that in table~\ref{tab:halfexceptional} and this is the simplest way to see intuitively how it generalises the patterns outlined in the introduction. 
We discuss this at the end of this section. 

First we look at the $\GL(d,\bbR)$ decompositions. Similarly to section~\ref{sec:spindd-decomp}, we use a different embedding to the one that immediately comes to mind. This leads to
\begin{equation}
	\adj(\SL(d+1,\bbR) \times \bbR^+)
		\simeq \bbR \oplus (T\otimes T^*) 
			\oplus (T\otimes \Lambda^d T) \oplus (T^* \otimes \Lambda^d T^*)
\end{equation}
We can identify here that the gauge field for the dual graviton living in $T^* \otimes \Lambda^d T^*$, which is not a pure differential form. Now let ($\rep{1}_{+1} \simeq \Lambda^d T^*$) and we have
\begin{equation}
	E \simeq ([1, 0, \dots , 0; 1] )_{+1}
		\simeq T \oplus (T^* \otimes \Lambda^{d-1} T^*)
			\oplus ((\Lambda^d T^*)^2 \otimes T^*)
\end{equation}
The dual graviton charge is thus $T^* \otimes \Lambda^{d-1} T^*$, and we also see here a higher dual charge $(\Lambda^d T^*)^2 \otimes T^*$, which must also result from pure gravity as that is all we have in this construction.

Now we examine
\begin{equation}
	S^2 E \simeq \rep{1}_{+2} \oplus [1, 0, \dots , 0; 1] _{+2}
		\oplus [0, 1, 0, \dots , 0, 1; 0] _{+2}
			 \oplus [2, 0, \dots , 0; 2] _{+2}
\end{equation}
and
\begin{equation}
	\Lambda^2 E \simeq [1, 0, \dots , 0; 1] _{+2}
		\oplus [2, 0, \dots , 0,1; 0] _{+2}
			 \oplus [0, 1, 0, \dots , 0; 2] _{+2}
\end{equation}
We see that $S^2 E$ has one term of the form $[2, 0, \dots , 0; *]$ while $\Lambda^2 E$ has a term like $[0,1, 0, \dots , 0; *]$. Applying the reasoning of~\ref{app:section} we have
\begin{equation}
	N \simeq \rep{1}_{+2} \oplus [1, 0, \dots , 0; 1] _{+2}
		\oplus [0, 1, 0, \dots , 0, 1; 0] _{+2} \subset S^2 E
\end{equation}
Looking at leading terms
\begin{equation}
\begin{aligned}[]
	[1, 0, \dots , 0; 1] _{+2} &\sim \Lambda^{d-1} T^* \oplus \dots \\
	[0, 1, 0, \dots , 0, 1; 0] _{+2} &\sim (T^*\otimes \Lambda^{d-2} T^*)^0 \oplus \dots
\end{aligned}
\end{equation}
we see that $N\simeq (T^*\otimes \Lambda^{d-2} T^*) \oplus (\Lambda^d T^*)^2 \oplus \dots$ looks to have the correct form under $\GL(d,\bbR)$ for there to be a gauge transformation of $E$ of the form
\begin{equation}
	\der : N \ra E
\end{equation}
as one would hope. However, this fails to be covariant as for the $d=7$ case of~\cite{CSW2}.

The natural guess for the ``torsion" representation is $E^* \oplus K \subset E^* \otimes \adj[\SL(d+1,\bbR) \times \bbR^+]$ where $K \sim \rep{1}_{-1} \oplus [1, 0, \dots , 0 ,1]_{-1} \oplus [0,1,0, \dots 0,1,0]_{-1}$. The decompositions are
\begin{equation}
\begin{aligned}[]
	[1, 0, \dots , 0 ,1]_{-1} &\simeq  T^* \oplus (T \otimes \Lambda^{d-1} T)
		\oplus ((\Lambda^d T)^2 \otimes T) \\
	[0, 1, 0, \dots , 0, 1, 0] _{-1} &\simeq (T \otimes \Lambda^2 T^*)^0 
		\oplus (T \otimes \Lambda^{d-1} T) \\
		&\qquad \oplus (\Lambda^d T \otimes \Lambda^2 T \otimes \Lambda^2 T^*)^0
		\oplus ((\Lambda^d T)^2 \otimes \Lambda^2 T \otimes T^*)^0
\end{aligned}
\end{equation}
so that the overall ``torsion" would be
\begin{equation}
\begin{aligned}[]
	E^* \oplus K \simeq \, & T^*\oplus (T \otimes \Lambda^2 T^*)
		\oplus 2 \times (T \otimes \Lambda^{d-1} T) \oplus \Lambda^d T
		\oplus (\Lambda^d T \otimes \Lambda^2 T \otimes \Lambda^2 T^*) \\
		&\qquad \oplus ((\Lambda^d T)^2 \otimes T )
		\oplus ((\Lambda^d T)^2 \otimes \Lambda^2 T \otimes T^*)
\end{aligned}
\end{equation}
However, even ignoring issues of covariance, the usual form~\eqref{eq:Dorfman-abs} of the Dorfman derivative and torsion~\eqref{eq:torsion-abs} fails to project out some parts of the connection. Therefore, it would seem that the prescription of~\cite{CSW1,CSW2,CSW3} would need some algebraic modification to include dual gravity.  An interesting approach to this modification can be found in~\cite{Hohm:2013jma}.

A final comment here is that the restriction to $d=D-2$ dimensions would involve the (centrally extended) affine algebra of~\cite{Geroch:1970nt,Geroch:1972yt} with diagram 
\begin{equation}
\label{eq:A+diagram}
\begin{tikzpicture} [place/.style={circle,draw=black,fill=white, inner sep=0pt,minimum size=8},baseline=(current  bounding  box.center)] 
\draw(-1,0)--(0,0); \draw(3,0)--(-2,1);
\draw (0,0)--(1,0);
\draw (2,0) -- (3,0); 
\draw[dashed] (1,0) -- (2,0);
\draw(-2,0)--(-1,0);
\draw (-2,0)--(-2,1);
\node at (1,0) [place,label=below:$$] {};
\node at (2,0) [place,label=below:$$] {}; 
\node at (3,0) [place,label=below:$(d-1)$] {};
\node at (0,0) [place,label=below:$$] {}; 
\node at (-1,0) [place,label=below:$2$] {}; 
\node at (-2,1) [place,label=left:$$,label=above:$d$] {}; 
\node at (-2,0) [place,label=left:$E$,label=below:$1$] {};
\ph{\node at (-4,0) [place,label=below:$$] {};}
\end{tikzpicture}
\end{equation}
This fits the pattern of the non-gravity line node connecting to those nodes of the gravity line corresponding to the potential term ($T^*\otimes \Lambda^{d-1} T^*$ in this case), though we do not wish to discuss further such infinite dimensional algebras here. 

The diagram~\eqref{eq:dual-grav-diagram} for the $d = D-3$ case can be viewed as 
that obtained by removing the node labelled $1$ from~\eqref{eq:A+diagram} (and relabelling $d\ra d+1$).
The node above the gravity line can then be thought of as connecting to the $T^*$ node but still keeping the $\Lambda^d T^*$ factor, which does not correspond to a node. This is why we positioned the added node over the left edge of the diagram.

This is similar to what happens in the simpler situation where there is a node for an $\SL(2,\bbR)$ factor added over the left-edge of the diagram corresponding to the addition of a $\Lambda^d T^*$ potential, such as in the $d=6$ case of table~\ref{tab:halfexceptional}. In these cases, there is a pattern that the Dynkin label of the generalised tangent space also has a 1 in the entry corresponding to the $\SL(2,\bbR)$ node.  
The exact same behaviour is seen in all series like that in table~\ref{tab:halfexceptional}, when one includes a new node for a top-form potential as the rank of the gravity line is increased. The $E_{3(3)} \simeq \SL(3,\bbR) \times\SL(2,\bbR)$ geometry in the exceptional series and the geometry of appendix~\ref{app:E77-IIA} provide further examples of this.


\subsection{Kaluza-Klein reduction of dual gravity}
\label{app:KK-dual-gravity}


Consider the setup of~\ref{app:dual-gravity} with $D=11$ and $d=8$, which is a subsector of eleven-dimensional supergravity restricted to eight-dimensions as we will see below. We wish to look at the reduction to type IIA, which corresponds to folding up the right-most node of the Dynkin diagram as in~\ref{app:KK}. This leads to the following $\GL(7,\bbR)$ decomposition for the remaining nodes of the gravity line
\begin{equation}
\begin{aligned}
	E \simeq T &\oplus \bbR \oplus \Lambda^6 T^*  \oplus \Lambda^7 T^*  \\ 
			&\oplus (\Lambda^7 T^*\otimes (\Lambda^7 T^*\oplus T^*)) \\
			& \oplus (T^* \otimes \Lambda^6 T^*) \oplus ((\Lambda^7 T^*)^2 \otimes T^* )
\end{aligned}
\end{equation}
In the usual type IIA language, we see terms for the D0 and D6-branes on the first line and the dual gravity setup of~\ref{app:dual-gravity}. The $\Lambda^7 T^*$ fits as the magnetic dual of the dilaton while the two terms on the middle line are higher duals of the D0 and D6-branes~\cite{Riccioni:2006az}.

In the usual gravity language, the D0-brane sources the Kaluza-Klein vector, while the D6-brane is its magnetic charge. The fact that the magnetic charge comes directly from this reduction makes sense of the idea that the starting setup includes a kind of magnetic dual of gravity.


\subsection{Some decompositions of $E_{8(8)} \times \bbR^+$}


Here we will briefly look at some of the other decompositions of $E_{8(8)} \times \bbR^+$, and their relevance to supergravity. We gave the $\GL(8,\bbR)$ decomposition and one of the $\Spin(8,8)$ decompositions in section~\ref{sec:tA6}. This subgroup resulted in a generalised geometry. The decompositions listed here do not lead to geometries (with the exception of~\ref{app:E77-IIA}), but are helpful examples for understanding how the Dynkin diagrams relate to the fields and charges. One sequence of embeddings we examine is
\begin{equation}
	E_{8(8)} \times \bbR^+ \ra O(8,8) \times\bbR^+ \ra O(7,7) \times O(1,1) \times \bbR^+
\end{equation}
where the $O(7,7)$ subgroup corresponds to the T-duality group of the type II theories. We emphasise that the middle group here is a different $O(8,8)$ subgroup to the one considered in section~\ref{sec:tA6}.

Another decomposition one can consider is that of $E_{8(8)} \times \bbR^+ \ra E_{7(7)} \times \SL(2,\bbR) \times \bbR^+$. The two orientations of the gravity line in this subalgebra will be seen to correspond to a subsector of the type IIB theory including the dual graviton and a new generalised geometry for a subsector of type IIA.


\subsubsection{$O(8,8)\times\bbR^+$ extension of T duality in 7 dimensions}
\label{app:O88T}

The (continuous) T-duality group in seven-dimensions is $O(7,7)$. This group is contained in an $O(8,8)$ subgroup of $E_{8(8)}$, so to begin with, we examine the decomposition $O(8,8)\times\bbR^+ \ra O(7,7) \times O(1,1) \times \bbR^+$
\begin{equation}
	\adj(O(8,8)\times\bbR^+)_{0}
		\ra \adj(O(7,7))_{(0,0)} \oplus \rep{7}_{(+1,0)} 
			\oplus \rep{7}_{(-1,0)} \oplus \rep{1}_{(0,0)}
\end{equation}
where in the pairs of weights on the right hand side, the first refers to the $O(1,1)$ factor, while the second refers to the original $\bbR^+$ factor in $O(8,8)\times\bbR^+$.

We now embed $\GL(7,\bbR)$ into $O(7,7) \times O(1,1) \times \bbR^+$ so that $\rep{1}_{(+1,0)} \simeq \Lambda^7 T$ and $\rep{1}_{(0,+1)} \simeq \Lambda^7 T^*$, and the embedding into the $O(7,7)$ factor is such that the vector decomposes as $T\oplus T^*$ as in~\cite{Gualtieri}. We then have the $\GL(7,\bbR)$ decompositions:
\begin{equation}
\begin{aligned}
	\rep{7}_{(+1,+1)} 
		&\simeq (T \oplus T^*) \otimes \Lambda^7 T \otimes \Lambda^7 T^* \\
		&\simeq T \oplus T^* \\
	\rep{7}_{(-1,+1)} 
		&\simeq (T \oplus T^*) \otimes ( \Lambda^7 T^*)^2 \\
		&\simeq (( \Lambda^7 T^*)^2\otimes T^*) 
			\oplus (\Lambda^7 T^*\otimes \Lambda^6 T^*) \\
	\rep{91}_{(0,+1)} 
		&\simeq ((T\otimes T^*) 
			\oplus \Lambda^2 T \oplus \Lambda^2 T^*) \otimes \Lambda^7 T^* \\
		&\simeq \Lambda^5 T^* \oplus (T^* \otimes \Lambda^6 T^*) 
			\oplus (\Lambda^7 T^*\otimes \Lambda^2 T^*) \\
	\rep{1}_{(0,+1)} &\simeq \Lambda^7 T^*
\end{aligned}
\end{equation}
The generalised tangent space for this group would be the adjoint of $O(8,8)$ with unit $\bbR^+$ weight.
\begin{equation}
\begin{aligned}
	E \simeq \rep{120}_{+1} 
		\simeq T &\oplus T^* \oplus \Lambda^5 T^*  
			\oplus (T^* \otimes \Lambda^6 T^*) \oplus \Lambda^7 T^* \\
			&\oplus (\Lambda^7 T^*\otimes \Lambda^6 T^*)
			\oplus (\Lambda^7 T^*\otimes \Lambda^2 T^*)
			\oplus ((\Lambda^7 T^*)^2 \otimes T^* )
\end{aligned}
\end{equation}
It is clear that the first line of this corresponds to the NS-NS sector complete with magnetic duals. The terms added to the usual tangent space correspond to the string, the NS5-brane, the dual graviton and the magnetic dual of the dilaton. The three terms on the second line are higher duals for the string, NS5-brane and graviton respectively.

Looking at the decomposition
\begin{equation}
\begin{aligned}
	E\otimes E 
		&\simeq [1, 0 , \dots , 0; 0,0]_{+1} \otimes [1, 0 , \dots , 0; 0,0]_{+1} \\
		& = \rep{1}_{+2} \oplus [0,0,1,0,0,0;0,0]_{+2} 
			\oplus [1,0,\dots,0;0,0]_{+2}
			 \oplus [0, \dots , 0; 0,2]_{+2} \\
			 	& \qquad \oplus [0,1,0,\dots, 0;0,1]_{+2} 
				\oplus [2,0 \dots, 0;0,0]_{+2}
\end{aligned}
\end{equation}
one can read off that
\begin{equation}
	N \simeq \rep{1}_{+2} \oplus [0, \dots , 0; 0,2]_{+2} \oplus [0,0,1,0,0,0;0,0]_{+2}
\end{equation}

The algebras for this magnetic completion of the NS-NS sector also form a series, whose diagrams we draw as
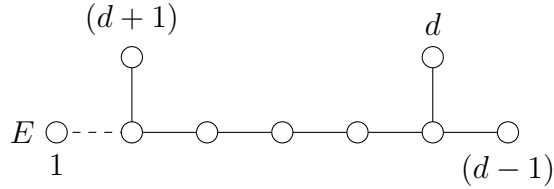
\begin{center}
\begin{tikzpicture} [place/.style={circle,draw=black,fill=white, inner sep=0pt,minimum size=8}] 
\draw(-1,0)--(0,0); \draw(-2,0)--(-2,1);
\draw (0,0)--(1,0);
\draw (2,0) -- (3,0); 
\draw (1,0) -- (2,0);
\draw(-2,0)--(-1,0);
\draw (2,0)--(2,1);
\draw[dashed](-3,0)--(-2,0);
\node at (1,0) [place,label=below:$$] {};
\node at (2,0) [place,label=below:$$] {}; 
\node at (2,1) [place,label=above:$d$] {}; 
\node at (3,0) [place,label=below:$(d-1)$] {};
\node at (0,0) [place,label=below:$$] {}; 
\node at (-1,0) [place,label=below:$$] {}; 
\node at (-2,1) [place,label=above:$(d+1)$] {}; 
\node at (-2,0) [place,label=left:$$] {};
\node at (-3,0) [place,label=left:$E$,label=below:$1$] {};
\ph{\node at (-4,0) [place,label=below:$$] {};}
\end{tikzpicture}
\end{center}
Due to the appearance of the potential for the dual graviton, these do not straightforwardly define generalised geometries in dimensions $d\geq7$. The case $d=6$ has group $\SO(6,6)\times\SL(2,\bbR)\times\bbR^+ \subset E_{7(7)} \times \bbR^+$ corresponding to a slight enhancement of ordinary generalised geometry by an $\SL(2,\bbR)$ factor. These algebras have been identified before~\cite{Schnakenburg:2004vd} in the very similar context of type I supergravity, and a similar algebra was considered in~\cite{west-conj} in the context of the bosonic string.


\subsubsection{$O(8,8)\times\bbR^+$ decompostion of $E_{8(8)} \times \bbR^+$}

Embedding the above in $E_{8(8)} \times \bbR^+$, gives us the $\GL(7,\bbR)$ decomposition of $E_{8(8)} \times \bbR^+$ relevant to the type II theories 
\begin{equation}
	E \simeq \rep{248}_{+1} \ra \adj(O(8,8))_{+1} \oplus \rep{128}^\pm_{+1}
\end{equation}
where we take $+$ for type IIB and $-$ for type IIA. The decomposition of the first term is the common NS-NS sector as above. The decomposition of the second term is
\begin{equation}
\begin{aligned}
	\rep{128}^+_{+1} \simeq \bbR &\oplus \Lambda^2 T^* 
		\oplus \Lambda^4 T^* \oplus \Lambda^6 T^* \\
		&\oplus \Big[ \Lambda^7 T^* \otimes \Big(
		\Lambda^7 T^* 
		\oplus \Lambda^5 T^* \oplus \Lambda^3 T^*
		\oplus T^*
		\Big) \Big] 
\end{aligned}
\end{equation}
for the type IIA case and
\begin{equation}
\begin{aligned}
	\rep{128}^-_{+1} \simeq T^* &\oplus \Lambda^3 T^* 
		\oplus \Lambda^5 T^* \oplus \Lambda^7 T^* \\
		&\oplus \Big[ \Lambda^7 T^* \otimes \Big(
		\Lambda^6 T^* 
		\oplus \Lambda^4 T^* \oplus \Lambda^2 T^*
		\oplus \bbR
		\Big) \Big] 
\end{aligned}
\end{equation}
for type IIB. These correspond to the D-branes of these theories, and their higher duals~\cite{Riccioni:2006az}.

As noted in~\cite{Schnakenburg:2001he,West:2004st}, the diagrams corresponding to the type IIA and IIB decompositions should be drawn as
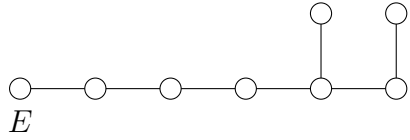
\begin{center}
\begin{tikzpicture} [place/.style={circle,draw=black,fill=white, inner sep=0pt,minimum size=8}] 
\draw(-1,0)--(0,0); \draw(2,0)--(2,1);
\draw (0,0)--(1,0);
\draw (1,0) -- (1,1); 
\draw (1,0) -- (2,0);
\draw(-2,0)--(-1,0);
\draw (-2,0)--(-3,0);
\node at (1,0) [place,label=below:$$] {};
\node at (2,0) [place,label=below:$$] {}; 
\node at (1,1) [place,label=below:$$] {};
\node at (0,0) [place,label=below:$$] {}; 
\node at (-1,0) [place,label=below:$$] {}; 
\node at (2,1) [place,label=right:$$] {}; 
\node at (-2,0) [place,label=below:$$] {};
\node at (-3,0) [place,label=below:$E$] {};
\end{tikzpicture}
\end{center}
for type IIA, the ``folding up" of the right-most node corresponding to KK reduction (as in~\ref{app:KK}), and
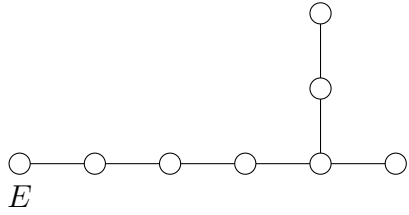
\begin{center}
\begin{tikzpicture} [place/.style={circle,draw=black,fill=white, inner sep=0pt,minimum size=8}] 
\draw(-1,0)--(0,0); \draw(1,0)--(1,1);
\draw (0,0)--(1,0);
\draw (1,1) -- (1,2); 
\draw (1,0) -- (2,0);
\draw(-2,0)--(-1,0);
\draw (-2,0)--(-3,0);
\node at (1,0) [place,label=below:$$] {};
\node at (2,0) [place,label=below:$$] {}; 
\node at (1,2) [place,label=below:$$] {};
\node at (0,0) [place,label=below:$$] {}; 
\node at (-1,0) [place,label=below:$$] {}; 
\node at (1,1) [place,label=right:$$] {}; 
\node at (-2,0) [place,label=below:$$] {};
\node at (-3,0) [place,label=below:$E$] {};
\end{tikzpicture}
\end{center}
for type IIB.


\subsubsection{$E_{7(7)} \times \SL(2,\bbR)$ in type IIB}
\label{app:E77-IIB}

Here we briefly look at the subsector corresponding to the $E_{7(7)} \times \SL(2,\bbR)\times\bbR^+$ subgroup in the type IIB decomposition. One has 
\begin{equation}
	\rep{248}_{+1} \ra \repp{133}{1}_{+1} \oplus \repp{56}{2}_{+1} + \repp{1}{3}_{+1}
\end{equation}
where
\begin{equation}
\begin{aligned}
	\repp{133}{1}_{+1} &\simeq T \oplus \Lambda^3 T^* 
		\oplus (T^* \otimes \Lambda^6 T^*) 
		\oplus (\Lambda^7 T^*\otimes \Lambda^4 T^*)
		\oplus ((\Lambda^7 T^*)^2 \otimes T^*) \\
	\repp{56}{2}_{+1} &\simeq 2 \times \Big[
		T^* \oplus \Lambda^5 T^* 
		\oplus (\Lambda^7 T^*\otimes \Lambda^2 T^*)
		\oplus (\Lambda^7 T^*\otimes \Lambda^6 T^*) \Big] \\
	\repp{1}{3}_{+1} &\simeq 3 \times (\Lambda^7 T^*)
\end{aligned}
\end{equation}
This corresponds to a different embedding of $\GL(7,\bbR)$ in $E_{7(7)}$ to the one considered in ~\cite{CSW2}. Clearly, this is a different reorganisation of the type IIB decomposition of the previous section, the $\SL(2,\bbR)$ factor corresponding to the S-duality symmetry. For this construction, we would take $E \simeq \repp{133}{1}_{+1}$ and draw the diagram
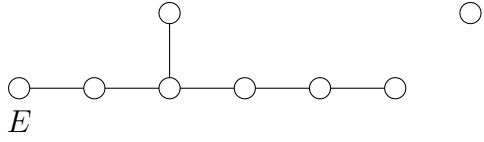
\begin{center}
\begin{tikzpicture} [place/.style={circle,draw=black,fill=white, inner sep=0pt,minimum size=8}] 
\draw(-1,0)--(0,0); \draw(0,0)--(0,1);
\draw (0,0)--(1,0);
\draw (2,0) -- (3,0); 
\draw (1,0) -- (2,0);
\draw(-2,0)--(-1,0);
\node at (1,0) [place,label=below:$$] {};
\node at (2,0) [place,label=below:$$] {}; 
\node at (3,0) [place,label=below:$$] {};
\node at (0,0) [place,label=below:$$] {}; 
\node at (-1,0) [place,label=below:$$] {}; 
\node at (0,1) [place,label=right:$$] {}; 
\node at (-2,0) [place,label=below:$E$] {};
\node at (4,1) [place,label=below:$$] {};
\end{tikzpicture}
\end{center}
the nodes added above the fourth node and zeroth node from the right indicating that the algebra is generated by $\Lambda^4 T \oplus \Lambda^4 T^*$ and $\Lambda^0 T \oplus \Lambda^0 T^*$. The interpretation of the very extended $E_7$ algebra as a subsector of type IIB has appeared before in~\cite{Kleinschmidt:2003mf} .

In this case one has
\begin{equation}
\begin{aligned}
	E\otimes E 
		\simeq \rep{1}_{+2} &\oplus [1,0,\dots,0,0;0]_{+2}
			\oplus [0,0,0,0,1,0;0]_{+2} \\
			 	& \oplus [0,1,0,\dots, 0;0]_{+2} 
				\oplus [2,0 \dots, 0;0]_{+2}
\end{aligned}
\end{equation}
so that
\begin{equation}
	N \simeq \rep{1}_{+2} \oplus [0,0,0,0,1,0;0]_{+2}
\end{equation}

Removing one node from the left of the diagram here, we recover the $d=6$ case of section~\ref{sec:spindd}, enhanced by an additional $\SL(2,\bbR)$ factor, which also includes the axion-dilaton system in the algebra.


\subsubsection{$E_{8(8)}$ adjoint in type IIB}

The adjoint of $E_{8(8)}$ is the $\rep{248}_0$ representation. We can get the $\GL(7,\bbR)$ decomposition of this from the $\rep{248}_{+1}$ above simply by multiplying all terms by $\Lambda^7 T$ to remove the $\bbR^+$ weight. This is typical of restrictions to $d=D-3$ dimensions, where the presence of the dual graviton typically makes the representation for $E$ a weighted version of the adjoint, and one can notice that the operation exchanges fields with their duals. This observation aids us in identifying the field relevant to each term. For brevity, we give only the type IIB decomposition
\begin{equation*}
\begin{aligned}
	\begin{pmatrix}
	\text{Graviton: } T \\
	\text{Dual Gravtion: } T^* \otimes \Lambda^6 T^* \\
	\text{2nd Dual: } ((\Lambda^7 T^*)^2 \otimes T^*)
	\end{pmatrix} &\stackrel{\otimes \Lambda^7 T}{\longrightarrow} ((T\otimes T^*) 
		\oplus (T\otimes \Lambda^7 T) \oplus (T^* \otimes \Lambda^7 T^*)) \\
	\begin{pmatrix}
	\text{F1: } T^* \\
	\text{Dual F1: } \Lambda^7 T^* \otimes \Lambda^6 T^*
	\end{pmatrix} &\longrightarrow (\Lambda^6 T \oplus \Lambda^6 T^*) \\
	\begin{pmatrix}
	\text{NS5: } \Lambda^5T^* \\
	\text{Dual NS5: } \Lambda^7 T^* \otimes \Lambda^2 T^*
	\end{pmatrix} &\longrightarrow (\Lambda^2 T \oplus \Lambda^2 T^*) \\
\end{aligned}
\end{equation*}
\begin{equation*}
\begin{aligned}
	\begin{pmatrix}
	\text{D1: } T^* \\
	\text{Dual D1: } \Lambda^7 T^* \otimes \Lambda^6 T^*
	\end{pmatrix} &\longrightarrow (\Lambda^6 T \oplus \Lambda^6 T^*) \\
	\begin{pmatrix}
	\text{D3: } \Lambda^3 T^* \\
	\text{Dual D3: } \Lambda^7 T^* \otimes \Lambda^4 T^*
	\end{pmatrix} &\longrightarrow (\Lambda^4 T \oplus \Lambda^4 T^*) \\
	\begin{pmatrix}
	\text{D5: } \Lambda^5 T^* \\
	\text{Dual D5: } \Lambda^7 T^* \otimes \Lambda^2 T^*
	\end{pmatrix} &\longrightarrow (\Lambda^2 T \oplus \Lambda^2 T^*) \\
	\begin{pmatrix}
	\text{D7: } \Lambda^7 T^* \\
	\text{Dual D7: } \Lambda^7 T^*
	\end{pmatrix} &\longrightarrow (\bbR \oplus \bbR) \\
	(\text{Dual dilaton: } \Lambda^7 T^* ) &\longrightarrow (\bbR)
\end{aligned}
\end{equation*}
The first part of the adjoint is the subsector of~\ref{app:dual-gravity}. The remaining parts exchange the fields and their duals, for example the parts of $E$ relevant to the NS5-brane becoming the parts of the adjoint relevant to the string and vice-versa. The remaining NS-NS seven-form in $E$ is mapped to the algebra generator $\bbR$ which corresponds to the dilaton, hence we identify this seven-form as the dual dilaton (as in~\ref{app:KK-dual-gravity}).


\subsubsection{$E_{7(7)} \times \SL(2,\bbR)$ in type IIA}
\label{app:E77-IIA}

Here we present a sketch of the subsector corresponding to the $E_{7(7)} \times \SL(2,\bbR)\times\bbR^+$ subgroup in the type IIA decomposition. This subgroup actually leads to a generalised geometry with diagram
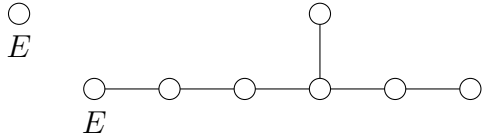
\begin{center}
\begin{tikzpicture} [place/.style={circle,draw=black,fill=white, inner sep=0pt,minimum size=8}] 
\draw(-1,0)--(0,0); \draw(1,0)--(1,1);
\draw (0,0)--(1,0);
\draw (2,0) -- (3,0); 
\draw (1,0) -- (2,0);
\draw(-2,0)--(-1,0);
\node at (1,0) [place,label=below:$$] {};
\node at (2,0) [place,label=below:$$] {}; 
\node at (3,0) [place,label=below:$$] {};
\node at (0,0) [place,label=below:$$] {}; 
\node at (-1,0) [place,label=below:$$] {}; 
\node at (1,1) [place,label=right:$$] {}; 
\node at (-2,0) [place,label=below:$E$] {};
\node at (-3,1) [place,label=below:$E$] {};
\end{tikzpicture}
\end{center}
This is the ``gravity-line-reversal" of the type IIB algebra of~\ref{app:E77-IIB}.

We choose the embedding of $\GL(7,\bbR)$ in $E_{7(7)} \times \SL(2,\bbR)\times\bbR^+$ such that
\begin{equation}
\begin{aligned}
	\adj (E_{7(7)}) &\simeq (T\otimes T^*) \oplus (\Lambda^3 T \oplus \Lambda^3 T^*)
		\oplus (\Lambda^6 T \oplus \Lambda^6 T^*) \\
	\adj (\SL(2,\bbR)) &\simeq \bbR \oplus \Lambda^7 T \oplus  \Lambda^7 T^* \\
	\rep{1}_{+1} &\simeq (\Lambda^7 T^*)
\end{aligned}
\end{equation}
The generalised tangent space corresponds to the $\repp{56}{2}_{+1}$ representation which then decomposes as
\begin{equation}
\begin{aligned}
	E \simeq T &\oplus \Lambda^2 T^* \oplus \Lambda^5 T^*  \oplus \Lambda^6 T^* \\
		&\oplus \Big[ \Lambda^7 T^* \otimes  
			 (\Lambda^5 T^* \oplus \Lambda^2 T^* \oplus  T^* ) \Big]  \\
		&\oplus ((\Lambda^7 T^*)^2 \otimes T^*)
\end{aligned}
\end{equation}
The first line includes the charges for the D2, NS5 and D6-branes, while the second line contains their dual charges. The last line is the higher dual of the graviton. The adjoint includes potentials only for the pure-form charges, so that there are no problems with the generalised geometric construction. It is an extension of the $E_{7(7)}\times\bbR^+$ geometry of~\cite{CSW2,CSW3} by the $\SL(2,\bbR)$ factor.

By the method of~\ref{app:section}, one finds that
\begin{equation}
	N \simeq \rep{1}_{+2} \oplus \repp{133}{2}_{+2} \oplus \repp{1539}{1}_{+2}
\end{equation}
while the only non-vanishing part of antisymmetric partial derivatives lives in the $\repp{1539}{3}_{+2}$ representation. 


\subsection{Six-dimensional $N=(1,0)$ supergravity}
\label{app:6D}

Here we briefly mention how one can see similar structures in six-dimensional theories with 8 supercharges. The corresponding infinite dimensional algebras and relation to the supergravity has previously appeared in~\cite{Riccioni:2008jz}.

The diagrams in this section feature a node for a short root added to the gravity line, in contrast to the other cases we have looked at, where all roots have the same length. This still adds the generators for the expected $p$-form potential into the system, but it will also lead to the inclusion of further terms which do not follow such a simple rule. 
A systematic discussion of this would be beyond the scope of the present paper, and we will content ourselves simply to note what happens in two simple examples.

For pure six-dimensional minimal supergravity restricted to three dimensions, consider $\SO(4,3)\times\bbR^+$ with the diagram\footnote{Note that this diagram is a collapsed version of the diagrams in section~\ref{app:O88T}.}
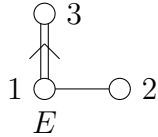
\begin{center}
\begin{tikzpicture} 
[place/.style={circle,draw=black, fill=white, inner sep=0pt,minimum size=8}] 
\draw (2.5,1) -- (1.5,1);\draw (1.45,1)--(1.45,2); \draw (1.55,1)--(1.55,2);
\draw (1.3,1.4)--(1.5,1.6); \draw(1.7,1.4)--(1.5,1.6);
\node at (2.5,1) [place,label=right:$2$] {};
\node at (1.5,1) [place,label=below:$E$,label=left:$1$] {};
\node at (1.5,2) [place,label=right:$3$] {};
\end{tikzpicture}
\end{center}
This leads to the $\GL(3,\bbR)$ decompositions
\begin{equation}
\begin{aligned}
	E &\simeq T \oplus T^* \oplus (T^* \otimes \Lambda^2 T^*) 
		\oplus (\Lambda^3 T^* \otimes \Lambda^2 T^*) 
			\oplus ((\Lambda^3T^*)^2 \otimes T^*) \\
	\adj &\simeq \bbR \oplus (T \otimes T^*) 
		\oplus \Lambda^2 T \oplus \Lambda^2 T^*
		\oplus (T \otimes \Lambda^3 T) \oplus (T^* \otimes \Lambda^3 T^*)
\end{aligned}
\end{equation}
showing that we have not only the expected two-form gauge field, but also the dual graviton. Due to the latter, the usual generalised geometry construction fails. The representation for $N$ comes out to be 
\begin{equation}
	N \simeq \rep{1}_{+2} \oplus [0,2;0]_{+2} \oplus [0,0;2]_{+2}
\end{equation}

One can see the exact same behaviour if one considers adding vector and tensor multiplets in six dimensions. The relevant group cosets are well-known (see~\cite{VanProeyen:2001wr} for a review of their geometry). For example, for the $\SO(5,4)\times\bbR^+$ case we draw the diagram as
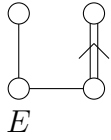
\begin{center}
\begin{tikzpicture} 
[place/.style={circle,draw=black, fill=white, inner sep=0pt,minimum size=8}] 
\draw (1.5,1) -- (1.5,2);
\draw (2.5,1) -- (1.5,1);\draw (2.45,1)--(2.45,2); \draw (2.55,1)--(2.55,2);
\draw (2.3,1.4)--(2.5,1.6); \draw(2.7,1.4)--(2.5,1.6);
\node at (2.5,1) [place,label=below:$$] {};
\node at (1.5,1) [place,label=below:$E$] {};
\node at (1.5,2) [place,label=right:$$] {};
\node at (2.5,2) [place,label=right:$$] {};
\end{tikzpicture}
\end{center}
and find the $\GL(3,\bbR)$ decompositions
\begin{equation}
\begin{aligned}
	E &\simeq T \oplus T^* \oplus (T^* \otimes \Lambda^2 T^*) 
		\oplus (\Lambda^3 T^* \otimes \Lambda^2 T^*) 
			\oplus ((\Lambda^3T^*)^2 \otimes T^*) \\
	& \qquad \oplus \Big[
		T^* \oplus \Lambda^3 T^* \oplus (\Lambda^3 T^* \otimes \Lambda^2 T^*) \Big] \\
	& \qquad \oplus \Big[ 
		\bbR \oplus \Lambda^2 T^* \oplus (\Lambda^3 T^* \otimes T^*)
		\oplus (\Lambda^3 T^*)^2 \Big] \\
	\adj &\simeq \bbR \oplus (T \otimes T^*) 
		\oplus \Lambda^2 T \oplus \Lambda^2 T^*
		\oplus (T \otimes \Lambda^3 T) \oplus (T^* \otimes \Lambda^3 T^*) \\
	& \qquad \oplus \Big[ \bbR \oplus \Lambda^2 T \oplus \Lambda^2 T^* \Big] \\
	& \qquad \oplus \Big[ T \oplus T^* \oplus \Lambda^3 T \oplus \Lambda^3 T^* \Big]
\end{aligned}
\end{equation}
which one can easily identify as adding a vector multiplet and a tensor multiplet (both with magnetic duals included) to the pure supergravity above. One can find similar decompositions for the theories related to the other very special quaternionic cosets.

The above algebras are relevant to restrictions of six-dimensional theories to $d=3$ dimensions. One could instead examine the restriction to $d=2$. In this case, there is no dual graviton and the construction of~\cite{CSW1,CSW2,CSW3} is expected to go through. However, this is of limited interest for studying backgrounds, as there can be no $H_{(3)}$ flux in two dimensions.




\begin{thebibliography}{99}



\bibitem{GCY}
   N.~Hitchin,
   ``Generalized Calabi-Yau manifolds,''
   Quart.\ J.\ Math.\ Oxford Ser.\  {\bf 54}, 281 (2003) 
   [arXiv:math.dg/0209099].
\bibitem{Gualtieri}
   M.~Gualtieri, 
   ``Generalized Complex Geometry,''
   Oxford University DPhil thesis (2004) [arXiv:math.DG/0401221] 
   and [arXiv:math.DG/0703298].
\bibitem{CSW1}
   A.~Coimbra, C.~Strickland-Constable, D.~Waldram,
   ``Supergravity as Generalised Geometry I: Type II Theories,''
  JHEP {\bf 1111} (2011) 091
  [arXiv:1107.1733 [hep-th]].
\bibitem{CSW2} 
   A.~Coimbra, C.~Strickland-Constable and D.~Waldram,
   ``$E_{d(d)} \times \mathbb{R}^+$ Generalised Geometry, Connections
   and M theory,'' 
  JHEP {\bf 1402}, 054 (2014)
  [arXiv:1112.3989 [hep-th].
\bibitem{CSW3} 
  A.~Coimbra, C.~Strickland-Constable and D.~Waldram,
  ``Supergravity as Generalised Geometry II: $E_{d(d)} \times \mathbb{R}^+$ and M theory,''
  JHEP {\bf 1403}, 019 (2014)
  [arXiv:1212.1586 [hep-th], arXiv:1212.1586].
\bibitem{Baraglia} 
  D.~Baraglia,
  ``Leibniz algebroids, twistings and exceptional generalized geometry,''
  J.\ Geom.\ Phys.\  {\bf 62}, 903 (2012)
  [arXiv:1101.0856 [math.DG]].

\bibitem{gen-susy1}
   M.~Grana, R.~Minasian, M.~Petrini and A.~Tomasiello,
   ``Supersymmetric backgrounds from generalized Calabi-Yau manifolds,''
   JHEP {\bf 0408} (2004) 046
   [arXiv:hep-th/0406137]. \\

\bibitem{gen-susy2}
   M.~Grana, R.~Minasian, M.~Petrini and A.~Tomasiello,
   ``Generalized structures of $N=1$ vacua,''
   JHEP {\bf 0511} (2005) 020
   [arXiv:hep-th/0505212].

\bibitem{gen-susy3}
   C.~Jeschek and F.~Witt,
   ``Generalised $G_2$-structures and type IIB superstrings,''
   JHEP {\bf 0503} (2005) 053
   [arXiv:hep-th/0412280].

\bibitem{gen-susy4} 
  P.~Berglund and P.~Mayr,
  ``Non-perturbative superpotentials in F-theory and string duality,''
  JHEP {\bf 1301}, 114 (2013)
  [JHEP {\bf 1301}, 114 (2013)]
  [hep-th/0504058].

\bibitem{sigmas1}
   U.~Lindstrom,
   ``Generalized N = (2,2) supersymmetric nonlinear sigma models,''
   Phys.\ Lett.\  B {\bf 587}, 216 (2004)
   [arXiv:hep-th/0401100],
\bibitem{sigmas2}
   U.~Lindstrom, R.~Minasian, A.~Tomasiello and M.~Zabzine,
   ``Generalized complex manifolds and supersymmetry,''
   Commun.\ Math.\ Phys.\  {\bf 257}, 235 (2005)
   [arXiv:hep-th/0405085],
\bibitem{sigmas3}
   A.~Kapustin,
   ``Topological strings on noncommutative manifolds,''
   Int.\ J.\ Geom.\ Meth.\ Mod.\ Phys.\  {\bf 1}, 49 (2004)
   [arXiv:hep-th/0310057],
\bibitem{sigmas4}
   A.~Kapustin and Y.~Li,
   ``Topological sigma-models with $H$-flux and twisted generalized complex
   manifolds,''
   arXiv:hep-th/0407249.

\bibitem{Gen-geom1} 
  M.~Grana, J.~Louis and D.~Waldram,
  ``Hitchin functionals in N=2 supergravity,''
  JHEP {\bf 0601}, 008 (2006)
  [hep-th/0505264].

\bibitem{Gen-geom2} 
  R.~Minasian, M.~Petrini and A.~Zaffaroni,
  ``Gravity duals to deformed SYM theories and Generalized Complex Geometry,''
  JHEP {\bf 0612}, 055 (2006)
  [hep-th/0606257].

\bibitem{Gen-geom3} 
  M.~Grana, J.~Louis and D.~Waldram,
  ``SU(3) x SU(3) compactification and mirror duals of magnetic fluxes,''
  JHEP {\bf 0704}, 101 (2007)
  [hep-th/0612237].
  
\bibitem{Gen-geom4} 
   I.~T.~Ellwood,
   ``NS-NS fluxes in Hitchin's generalized geometry,''
   JHEP {\bf 0712}, 084 (2007)
   [arXiv:hep-th/0612100].

\bibitem{Gen-geom5} 
  P.~Koerber and L.~Martucci,
  ``From ten to four and back again: How to generalize the geometry,''
  JHEP {\bf 0708}, 059 (2007)
  [arXiv:0707.1038 [hep-th]].

\bibitem{Gen-geom6} 
  D.~Cassani and A.~Bilal,
  ``Effective actions and N=1 vacuum conditions from SU(3) x SU(3) compactifications,''
  JHEP {\bf 0709}, 076 (2007)
  [arXiv:0707.3125 [hep-th]].

\bibitem{Gen-geom7} 
  D.~Cassani,
  ``Reducing democratic type II supergravity on SU(3) x SU(3) structures,''
  JHEP {\bf 0806}, 027 (2008)
  [arXiv:0804.0595 [hep-th]].

\bibitem{Gen-geom8} 
   M.~Grana, R.~Minasian, M.~Petrini and D.~Waldram,
   ``T-duality, Generalized Geometry and Non-Geometric Backgrounds,''
   JHEP {\bf 0904}, 075 (2009)
   [arXiv:0807.4527 [hep-th]].

\bibitem{Gen-geom9} 
  M.~Gabella, J.~P.~Gauntlett, E.~Palti, J.~Sparks and D.~Waldram,
  ``AdS(5) Solutions of Type IIB Supergravity and Generalized Complex Geometry,''
  Commun.\ Math.\ Phys.\  {\bf 299}, 365 (2010)
  [arXiv:0906.4109 [hep-th]].

\bibitem{Gen-geom10} 
  A.~Tomasiello,
  ``Generalized structures of ten-dimensional supersymmetric solutions,''
  JHEP {\bf 1203}, 073 (2012)
  [arXiv:1109.2603 [hep-th]].

\bibitem{Gen-geom11} 
  M.~Petrini and A.~Zaffaroni,
  ``A Note on Supersymmetric Type II Solutions of Lifshitz Type,''
  JHEP {\bf 1207}, 051 (2012)
  [arXiv:1202.5542 [hep-th]].

\bibitem{Gen-geom12} 
  A.~Kahle and R.~Minasian,
  ``D-brane couplings and Generalised Geometry,''
  arXiv:1301.7238 [hep-th].

\bibitem{Gen-geom13} 
  D.~Rosa and A.~Tomasiello,
  ``Pure spinor equations to lift gauged supergravity,''
  JHEP {\bf 1401}, 176 (2014)
  [arXiv:1305.5255 [hep-th]].

\bibitem{Gen-geom14} 
  D.~Prins and D.~Tsimpis,
  ``IIB supergravity on manifolds with SU(4) structure and generalized geometry,''
  JHEP {\bf 1307}, 180 (2013)
  [arXiv:1306.2543 [hep-th]].

\bibitem{Gen-geom15} 
  D.~Andriot and A.~Betz,
  ``$\beta$-supergravity: a ten-dimensional theory with non-geometric fluxes, and its geometric framework,''
  JHEP {\bf 1312}, 083 (2013)
  [arXiv:1306.4381 [hep-th]].

\bibitem{gen-calib1} 
  P.~Koerber,
  ``Stable D-branes, calibrations and generalized Calabi-Yau geometry,''
  JHEP {\bf 0508}, 099 (2005)
  [hep-th/0506154].

\bibitem{gen-calib2} 
  L.~Martucci and P.~Smyth,
  ``Supersymmetric D-branes and calibrations on general N=1 backgrounds,''
  JHEP {\bf 0511}, 048 (2005)
  [hep-th/0507099].

\bibitem{gen-calib3} 
  L.~Martucci,
  ``D-branes on general N=1 backgrounds: Superpotentials and D-terms,''
  JHEP {\bf 0606}, 033 (2006)
  [hep-th/0602129].

\bibitem{gen-calib4} 
  P.~Koerber and L.~Martucci,
  ``Deformations of calibrated D-branes in flux generalized complex manifolds,''
  JHEP {\bf 0612}, 062 (2006)
  [hep-th/0610044].

\bibitem{gen-calib5} 
  D.~Lust, F.~Marchesano, L.~Martucci and D.~Tsimpis,
  ``Generalized non-supersymmetric flux vacua,''
  JHEP {\bf 0811}, 021 (2008)
  [arXiv:0807.4540 [hep-th]].

\bibitem{Koerber:2010bx} 
  P.~Koerber,
  ``Lectures on Generalized Complex Geometry for Physicists,''
  Fortsch.\ Phys.\  {\bf 59}, 169 (2011)
  [arXiv:1006.1536 [hep-th]].


\bibitem{Hull07}
   C.~M.~Hull,
   ``Generalised Geometry for M-Theory,''
   JHEP {\bf 0707}, 079 (2007)
   [arXiv:hep-th/0701203].

\bibitem{PW}
   P.~P.~Pacheco, D.~Waldram,
   ``M-theory, exceptional generalised geometry and superpotentials,''
   JHEP {\bf 0809}, 123 (2008).
   [arXiv:0804.1362 [hep-th]].

\bibitem{Triendl:2009ap} 
  H.~Triendl and J.~Louis,
  ``Type II compactifications on manifolds with SU(2) x SU(2) structure,''
  JHEP {\bf 0907}, 080 (2009)
  [arXiv:0904.2993 [hep-th]].

\bibitem{GLSW}
   M.~Grana, J.~Louis, A.~Sim, D.~Waldram,
   ``$E_{7(7)}$ formulation of $N=2$ backgrounds,''
   JHEP {\bf 0907}, 104 (2009).
   [arXiv:0904.2333 [hep-th]].

\bibitem{E7-flux}
   G.~Aldazabal, E.~Andres, P.~G.~Camara and M.~Grana,
   ``U-dual fluxes and Generalized Geometry,''
   JHEP\ {\bf 1011}, 083  (2010)
   [arXiv:1007.5509 [hep-th]].

\bibitem{GO1}
   M.~Grana and F.~Orsi,
   ``$N=1$ vacua in Exceptional Generalized Geometry,''
   JHEP {\bf 1108}, 109 (2011)
   [arXiv:1105.4855 [hep-th]].

\bibitem{GO2}
   M.~Grana and F.~Orsi,
   ``$N=2$ vacua in Generalized Geometry,''
   JHEP {\bf 1211}, 052 (2012)
   [arXiv:1207.3004 [hep-th]].

\bibitem{GT} 
   M.~Grana and H.~Triendl,
   ``Generalized $N=1$ and $N=2$ structures in M-theory and type II
   orientifolds,'' 
  JHEP {\bf 1303}, 145 (2013)
  [arXiv:1211.3867 [hep-th]].

\bibitem{siegel1}
   W.~Siegel,
   ``Two vierbein formalism for string inspired axionic gravity,''
   Phys.\ Rev.\  {\bf D47 } (1993)  5453-5459.
   [hep-th/9302036],
\bibitem{siegel2}
   W.~Siegel,
   ``Superspace duality in low-energy superstrings,''
   Phys.\ Rev.\  {\bf D48 } (1993)  2826-2837.
   [hep-th/9305073].

\bibitem{Tfold}
   C.~M.~Hull,
   ``A geometry for non-geometric string backgrounds,''
   JHEP {\bf 0510}, 065 (2005)
   [arXiv:hep-th/0406102].

\bibitem{dft}
   C.~Hull, B.~Zwiebach,
   ``Double Field Theory,''
   JHEP {\bf 0909 } (2009)  099.
   [arXiv:0904.4664 [hep-th]].

\bibitem{Hohm:2010jy} 
  O.~Hohm, C.~Hull and B.~Zwiebach,
  ``Background independent action for double field theory,''
  JHEP {\bf 1007}, 016 (2010)
  [arXiv:1003.5027 [hep-th]].

\bibitem{DFT1}
  O.~Hohm, C.~Hull, B.~Zwiebach,
  ``Generalized metric formulation of double field theory,''
  JHEP {\bf 1008 } (2010)  008.
  [arXiv:1006.4823 [hep-th]].

\bibitem{DFT2}
  O.~Hohm, S.~K.~Kwak,
  ``Frame-like Geometry of Double Field Theory,''
  J.\ Phys.\ A {\bf A44 } (2011)  085404.
  [arXiv:1011.4101 [hep-th]].

\bibitem{DFT3}
  I.~Jeon, K.~Lee, J.~-H.~Park,
  ``Differential geometry with a projection: Application to double
  field theory,'' 
  JHEP {\bf 1104 } (2011)  014.
  [arXiv:1011.1324 [hep-th]],
\bibitem{DFT3b}
  I.~Jeon, K.~Lee, J.~-H.~Park,
  ``Stringy differential geometry, beyond Riemann,''  
  Phys.\ Rev.\ D {\bf 84}, 044022 (2011)
  [arXiv:1105.6294 [hep-th]].

\bibitem{DFT4}
  O.~Hohm, S.~K.~Kwak, B.~Zwiebach,
  ``Double Field Theory of Type II Strings,''
  JHEP {\bf 1109}, 013 (2011)
  [arXiv:1107.0008 [hep-th]].

\bibitem{DFT5}
  G.~Aldazabal, W.~Baron, D.~Marques and C.~Nunez,
  ``The effective action of Double Field Theory,''
  JHEP {\bf 1111}, 052 (2011)
  [Erratum-ibid.\  {\bf 1111}, 109 (2011)]
  [arXiv:1109.0290 [hep-th]].

\bibitem{DFT6}
  I.~Jeon, K.~Lee, J.~-H.~Park and Y.~Suh,
  ``Stringy Unification of Type IIA and IIB Supergravities under N=2 D=10 Supersymmetric Double Field Theory,''
  Phys.\ Lett.\ B {\bf 723}, 245 (2013)
  [arXiv:1210.5078 [hep-th]].

\bibitem{DFT7}
  O.~Hohm and B.~Zwiebach,
  ``Towards an invariant geometry of double field theory,''
  J.\ Math.\ Phys.\  {\bf 54}, 032303 (2013)
  [arXiv:1212.1736 [hep-th]].

\bibitem{DFT8}
  D.~S.~Berman and K.~Lee,
  ``Supersymmetry for Gauged Double Field Theory and Generalised Scherk-Schwarz Reductions,''
  Nucl.\ Phys.\ B {\bf 881}, 369 (2014)
  [arXiv:1305.2747 [hep-th]].

\bibitem{DFT9}
  O.~Hohm, W.~Siegel and B.~Zwiebach,
  ``Doubled $\alpha'$-Geometry,''
  JHEP {\bf 1402}, 065 (2014)
  [arXiv:1306.2970 [hep-th]].

\bibitem{DFT10}
  O.~Hohm and H.~Samtleben,
  ``Gauge theory of Kaluza-Klein and winding modes,''
  Phys.\ Rev.\ D {\bf 88}, 085005 (2013)
  [arXiv:1307.0039 [hep-th]].

\bibitem{DFT11}
  L.~Freidel, R.~G.~Leigh and D.~Minic,
  ``Born Reciprocity in String Theory and the Nature of Spacetime,''
  Phys.\ Lett.\ B {\bf 730}, 302 (2014)
  [arXiv:1307.7080 [hep-th]].

\bibitem{DFT12}
  G.~Aldazabal, D.~Marques and C.~Nunez,
  ``Double Field Theory: A Pedagogical Review,''
  Class.\ Quant.\ Grav.\  {\bf 30}, 163001 (2013)
  [arXiv:1305.1907 [hep-th]].

\bibitem{DFT13}
  O.~Hohm, D.~Lust and B.~Zwiebach,
  ``The Spacetime of Double Field Theory: Review, Remarks, and Outlook,''
  Fortsch.\ Phys.\  {\bf 61}, 926 (2013)
  [arXiv:1309.2977 [hep-th]].

\bibitem{BP1}
   D.~S.~Berman and M.~J.~Perry,
   ``Generalized Geometry and M theory,''
   JHEP\ {\bf 1106}, 074  (2011)
   [arXiv:1008.1763 [hep-th]]; \\

\bibitem{BP2}
   D.~S.~Berman, H.~Godazgar and M.~J.~Perry,
   ``$SO(5,5)$ duality in M-theory and generalized geometry,''
   Phys.\ Lett.\ B\ {\bf 700}, 65  (2011)
   [arXiv:1103.5733 [hep-th]]; \\

\bibitem{BP-alg}
   D.~S.~Berman, H.~Godazgar, M.~Godazgar and M.~J.~Perry,
   ``The Local symmetries of M-theory and their formulation in
   generalised geometry,'' 
   JHEP {\bf 1201}, 012 (2012)
   [arXiv:1110.3930 [hep-th]].

\bibitem{BPW}
   D.~S.~Berman, H.~Godazgar, M.~J.~Perry and P.~West,
   ``Duality Invariant Actions and Generalised Geometry,''
   JHEP {\bf 1202}, 108 (2012)
   [arXiv:1111.0459 [hep-th]].

\bibitem{BCKT}
   D.~S.~Berman, M.~Cederwall, A.~Kleinschmidt and D.~C.~Thompson,
   ``The gauge structure of generalised diffeomorphisms,''
  JHEP {\bf 1301}, 064 (2013)
  [arXiv:1208.5884 [hep-th]].

\bibitem{MTheoryDFT1} 
   D.~S.~Berman, E.~T.~Musaev and D.~C.~Thompson,
   ``Duality Invariant M-theory: Gauged supergravities and
   Scherk-Schwarz reductions,'' 
   JHEP {\bf 1210}, 174 (2012)
   [arXiv:1208.0020 [hep-th]].

\bibitem{MTheoryDFT2} 
  E.~T.~Musaev,
  ``Gauged supergravities in 5 and 6 dimensions from generalised Scherk-Schwarz reductions,''
  JHEP {\bf 1305}, 161 (2013)
  [arXiv:1301.0467 [hep-th]].

\bibitem{MTheoryDFT3} 
  G.~Aldazabal, M.~Gra–a, D.~MarquŽs and J.~A.~Rosabal,
  ``Extended geometry and gauged maximal supergravity,''
  JHEP {\bf 1306}, 046 (2013)
  [arXiv:1302.5419 [hep-th]].

\bibitem{MTheoryDFT4} 
  M.~Cederwall, J.~Edlund and A.~Karlsson,
  ``Exceptional geometry and tensor fields,''
  JHEP {\bf 1307}, 028 (2013)
  [arXiv:1302.6736 [hep-th]].

\bibitem{MTheoryDFT5} 
  O.~Hohm and H.~Samtleben,
  ``Exceptional Form of D=11 Supergravity,''
  Phys.\ Rev.\ Lett.\  {\bf 111}, 231601 (2013)
  [arXiv:1308.1673 [hep-th]].

\bibitem{MTheoryDFT6} 
  D.~S.~Berman and D.~C.~Thompson,
  ``Duality Symmetric String and M-Theory,''
  Phys.\ Rept.\  {\bf 566}, 1 (2014)
  [arXiv:1306.2643 [hep-th]].

\bibitem{Andriot} 
  D.~Andriot,
  ``Heterotic string from a higher dimensional perspective,''
  Nucl.\ Phys.\ B {\bf 855}, 222 (2012)
  [arXiv:1102.1434 [hep-th]].

\bibitem{BnGeom}
   R.~Rubio,
   ``$B_n$-Generalized geometry and $G^2_2$-structures,"
   [arXiv:1301.3330 [math.DG]].

\bibitem{Hohm:2011ex} 
  O.~Hohm and S.~K.~Kwak,
  ``Double Field Theory Formulation of Heterotic Strings,''
  JHEP {\bf 1106}, 096 (2011)
  [arXiv:1103.2136 [hep-th]].

\bibitem{Hohm:2011nu} 
  O.~Hohm and S.~K.~Kwak,
  ``N=1 Supersymmetric Double Field Theory,''
  JHEP {\bf 1203}, 080 (2012)
  [arXiv:1111.7293 [hep-th]].

\bibitem{courant-reduction}
   H.~Bursztyn, G.~Cavalcanti and M.~Gualtieri, 
   ``Reduction of Courant algebroids and generalized complex structures," 
   Adv. Math. {\bf 211} (2) (2007) 
   [arxiv:0509.640 [math.DG]].

\bibitem{heterotic1}
   M.~Garcia-Fernandez,
   ``Torsion-free generalized connections and Heterotic Supergravity,"
  Commun.\ Math.\ Phys.\  {\bf 332}, no. 1, 89 (2014)
  [arXiv:1304.4294 [math.DG]].
\bibitem{heterotic2}
   D.~Baraglia and P.~Hekmati
   ``Transitive Courant Algebroids, String Structures and T-duality,"
  Adv.\ Theor.\ Math.\ Phys.\  {\bf 19}, 613 (2015)
  [arXiv:1308.5159 [math.DG]].

\bibitem{Kleinschmidt:2003mf} 
  A.~Kleinschmidt, I.~Schnakenburg and P.~C.~West,
  ``Very extended Kac-Moody algebras and their interpretation at low levels,''
  Class.\ Quant.\ Grav.\  {\bf 21}, 2493 (2004)
  [hep-th/0309198].


\bibitem{Curtright:1980yk} 
  T.~Curtright,
  ``Generalized Gauge Fields,''
  Phys.\ Lett.\ B {\bf 165}, 304 (1985).

\bibitem{west-conj} 
   P.~C.~West,
   ``$E_{11}$ and M theory,''
   Class.\ Quant.\ Grav.\ \ {\bf 18}, 4443  (2001)
   [hep-th/0104081]; \\

\bibitem{Hull:2001iu} 
  C.~M.~Hull,
  ``Duality in gravity and higher spin gauge fields,''
  JHEP {\bf 0109}, 027 (2001)
  [hep-th/0107149].

\bibitem{West:2012qm} 
  P.~West,
  ``Generalised BPS conditions,''
  Mod.\ Phys.\ Lett.\ A {\bf 27}, 1250202 (2012)
  [arXiv:1208.3397 [hep-th]].


\bibitem{Hohm:2013jma} 
  O.~Hohm and H.~Samtleben,
  ``U-duality covariant gravity,''
  JHEP {\bf 1309}, 080 (2013)
  [arXiv:1307.0509 [hep-th]].

\bibitem{deWit:2008ta} 
  B.~de Wit, H.~Nicolai and H.~Samtleben,
  ``Gauged Supergravities, Tensor Hierarchies, and M-Theory,''
  JHEP {\bf 0802}, 044 (2008)
  [arXiv:0801.1294 [hep-th]].

\bibitem{West:2002jj} 
  P.~C.~West,
  ``Very extended E(8) and A(8) at low levels, gravity and supergravity,''
  Class.\ Quant.\ Grav.\  {\bf 20}, 2393 (2003)
  [hep-th/0212291].



\bibitem{MixedSym1}
  P.~de Medeiros and C.~Hull,
  ``Exotic tensor gauge theory and duality,''
  Commun.\ Math.\ Phys.\  {\bf 235}, 255 (2003)
  [hep-th/0208155].

\bibitem{MixedSym2}
  X.~Bekaert, N.~Boulanger and M.~Henneaux,
  ``Consistent deformations of dual formulations of linearized gravity: A No go result,''
  Phys.\ Rev.\ D {\bf 67}, 044010 (2003)
  [hep-th/0210278].

\bibitem{MixedSym3}
  P.~de Medeiros and C.~Hull,
  ``Geometric second order field equations for general tensor gauge fields,''
  JHEP {\bf 0305}, 019 (2003)
  [hep-th/0303036].

\bibitem{MixedSym4}
  M.~Henneaux and C.~Teitelboim,
  ``Duality in linearized gravity,''
  Phys.\ Rev.\ D {\bf 71}, 024018 (2005)
  [gr-qc/0408101].

\bibitem{MixedSym5}
  N.~Boulanger and O.~Hohm,
  ``Non-linear parent action and dual gravity,''
  Phys.\ Rev.\ D {\bf 78}, 064027 (2008)
  [arXiv:0806.2775 [hep-th]].

\bibitem{MixedSym6}
  M.~Henneaux, A.~Kleinschmidt and H.~Nicolai,
  ``Real forms of extended Kac-Moody symmetries and higher spin gauge theories,''
  Gen.\ Rel.\ Grav.\  {\bf 44}, 1787 (2012)
  [arXiv:1110.4460 [hep-th]].

\bibitem{MixedSym7}
  P.~West,
  ``Generalised geometry, eleven dimensions and E11,''
  JHEP {\bf 1202}, 018 (2012)
  [arXiv:1111.1642 [hep-th]].

\bibitem{MixedSym8}
  H.~Godazgar, M.~Godazgar and M.~J.~Perry,
  ``E8 duality and dual gravity,''
  JHEP {\bf 1306}, 044 (2013)
  [arXiv:1303.2035 [hep-th]].

\bibitem{MixedSym9}
  C.~Bunster and M.~Henneaux,
  ``Sources for Generalized Gauge Fields,''
  Phys.\ Rev.\ D {\bf 88}, 085002 (2013)
  [arXiv:1308.2866 [hep-th]].

\bibitem{MixedSym10}
  P.~P.~Cook and M.~Fleming,
  ``Gravitational Coset Models,''
  JHEP {\bf 1407}, 115 (2014)
  [arXiv:1309.0757 [hep-th]].

\bibitem{MixedSym11}
  H.~Godazgar, M.~Godazgar and H.~Nicolai,
  ``Generalised geometry from the ground up,''
  JHEP {\bf 1402}, 075 (2014)
  [arXiv:1307.8295 [hep-th]].

\bibitem{MixedSym12}
  H.~Godazgar, M.~Godazgar and H.~Nicolai,
  ``Non-linear Kaluza-Klein theory for dual fields,''
  Phys.\ Rev.\ D {\bf 88}, no. 12, 125002 (2013)
  [arXiv:1309.0266 [hep-th]].

\bibitem{HT}
   C.~M.~Hull and P.~K.~Townsend,
   ``Unity of superstring dualities,''
   Nucl.\ Phys.\  B {\bf 438} (1995) 109
   [arXiv:hep-th/9410167].

\bibitem{Julia1}
  E.~Cremmer and B.~Julia,
  ``The SO(8) Supergravity,''
  Nucl.\ Phys.\ B {\bf 159}, 141 (1979).

\bibitem{Julia2}
  B.~Julia,
  ``Infinite Lie Algebras In Physics,''
  In *Baltimore 1981, Proceedings, Unified Field Theories and Beyond*, 23-41

\bibitem{Julia3}
  B.~Julia,
  ``Kac-moody Symmetry Of Gravitation And Supergravity Theories,''
  LPTENS-82-22.

\bibitem{deWN1}
   B.~de Wit and H.~Nicolai,
   ``$D = 11$ Supergravity With Local $\SU(8)$ Invariance,''
   Nucl.\ Phys.\ B {\bf 274}, 363 (1986).

\bibitem{deWN2}
   H.~Nicolai,
   ``$D = 11$ Supergravity with Local $\SO(16)$ Invariance,''
   Phys.\ Lett.\  B {\bf 187}, 316 (1987).

\bibitem{deWN3}
   M.~J.~Duff,
   ``$E_8\times\SO(16)$ Symmetry of $d=11$ Supergravity,''
   in \textit{Quantum field theory and quantum statistics}, vol. 2,
   p209, eds. I.~A.~Batalin et al., Adam Hilger (1987) (CERN-TH-4124). 

\bibitem{deWN4}
  H.~Nicolai and N.~P.~Warner,
  ``The Structure of $N=16$ Supergravity in Two-dimensions,''
  Commun.\ Math.\ Phys.\  {\bf 125}, 369 (1989).

\bibitem{deWN5}
  H.~Nicolai,
  ``A Hyperbolic Lie algebra from supergravity,''
  Phys.\ Lett.\ B {\bf 276}, 333 (1992).

\bibitem{deWN6}
   K.~Koepsell, H.~Nicolai and H.~Samtleben,
   ``An exceptional geometry for $d=11$ supergravity?,''
   Class.\ Quant.\ Grav.\  {\bf 17}, 3689 (2000)
   [arXiv:hep-th/0006034].

\bibitem{deWN7}
   B.~de Wit,
   ``M theory duality and BPS extended supergravity,''
   Int.\ J.\ Mod.\ Phys.\ A {\bf 16}, 1002 (2001)
   [hep-th/0010292].

\bibitem{deWN8}
   B.~de Wit and H.~Nicolai,
   ``Hidden symmetries, central charges and all that,''
   Class.\ Quant.\ Grav.\ \ {\bf 18}, 3095  (2001)
   [hep-th/0011239].

\bibitem{deWN9}
   C.~Hillmann,
   ``Generalized $E_{7(7)}$ coset dynamics and $D=11$ supergravity,''
   JHEP {\bf 0903}, 135 (2009).
   [arXiv:0901.1581 [hep-th]], 
\bibitem{deWN10}
   C.~Hillmann,
   ``$E_{7(7)}$ and $d=11$ supergravity,''
   [arXiv:0902.1509 [hep-th]].

\bibitem{west1} 
   P.~C.~West,
   ``Hidden superconformal symmetry in M theory,''
   JHEP {\bf 0008}, 007 (2000)
   [hep-th/0005270]; 
\bibitem{west2} 
   P.~C.~West,
   ``E(11), SL(32) and central charges,''
   Phys.\ Lett.\ B {\bf 575}, 333 (2003)
   [hep-th/0307098].

\bibitem{E10-1}
   T.~Damour, M.~Henneaux and H.~Nicolai,
   ``$E_{10}$ and a 'small tension expansion' of M theory,''
   Phys.\ Rev.\ Lett.\  {\bf 89}, 221601 (2002)
   [hep-th/0207267].
\bibitem{E10-2}
   T.~Damour, M.~Henneaux and H.~Nicolai,
   ``Cosmological billiards,''
   Class.\ Quant.\ Grav.\  {\bf 20}, R145 (2003)
   [hep-th/0212256].


\bibitem{Cremmer:1999du} 
  E.~Cremmer, B.~Julia, H.~Lu and C.~N.~Pope,
  ``Higher dimensional origin of D = 3 coset symmetries,''
  hep-th/9909099.

\bibitem{Nicolai:2003fw} 
  H.~Nicolai and T.~Fischbacher,
  ``Low level representations for E(10) and E(11),''
  hep-th/0301017.

\bibitem{Riccioni:2006az} 
  F.~Riccioni and P.~C.~West,
  ``Dual fields and E(11),''
  Phys.\ Lett.\ B {\bf 645}, 286 (2007)
  [hep-th/0612001].

\bibitem{Schnakenburg:2001he} 
  I.~Schnakenburg and P.~C.~West,
  ``Kac-Moody symmetries of 2B supergravity,''
  Phys.\ Lett.\ B {\bf 517}, 421 (2001)
  [hep-th/0107181].

\bibitem{West:2004st} 
  P.~C.~West,
  ``The IIA, IIB and eleven-dimensional theories and their common E(11) origin,''
  Nucl.\ Phys.\ B {\bf 693}, 76 (2004)
  [hep-th/0402140].

\bibitem{Schnakenburg:2004vd} 
  I.~Schnakenburg and P.~C.~West,
  ``Kac-Moody symmetries of ten-dimensional nonmaximal supergravity theories,''
  JHEP {\bf 0405}, 019 (2004)
  [hep-th/0401196].

\bibitem{Riccioni:2008jz} 
  F.~Riccioni, A.~Van Proeyen and P.~C.~West,
  ``Real forms of very extended Kac-Moody algebras and theories with eight supersymmetries,''
  JHEP {\bf 0805}, 079 (2008)
  [arXiv:0801.2763 [hep-th]].

\bibitem{Houart:2009ed} 
  L.~Houart, A.~Kleinschmidt, J.~Lindman Hornlund, D.~Persson and N.~Tabti,
  ``Finite and infinite-dimensional symmetries of pure N=2 supergravity in D=4,''
  JHEP {\bf 0908}, 098 (2009)
  [arXiv:0905.4651 [hep-th]].

\bibitem{CSW4}
   A.~Coimbra, C.~Strickland-Constable, D.~Waldram,
  ``Supersymmetric Backgrounds and Generalised Special Holonomy,''
  Class.\ Quant.\ Grav.\  {\bf 33}, no. 12, 125026 (2016)
  [arXiv:1411.5721 [hep-th]].

\bibitem{MicuTalk}
   A.~Micu,
   Talk given at String Phenomenology 2013, DESY, Hamburg

\bibitem{West:2010ev} 
  P.~West,
  ``$E_{11}$, generalised space-time and IIA string theory,''
  Phys.\ Lett.\ B {\bf 696}, 403 (2011)
  [arXiv:1009.2624 [hep-th]].



\bibitem{7Branes1}
  P.~Meessen and T.~Ortin,
  ``An Sl(2,Z) multiplet of nine-dimensional type II supergravity theories,''
  Nucl.\ Phys.\ B {\bf 541}, 195 (1999)
  [hep-th/9806120].

\bibitem{7Branes2}
  G.~Dall'Agata, K.~Lechner and M.~Tonin,
  ``D = 10, N = IIB supergravity: Lorentz invariant actions and duality,''
  JHEP {\bf 9807}, 017 (1998)
  [hep-th/9806140].

\bibitem{Geroch:1970nt} 
  R.~P.~Geroch,
  ``A Method for generating solutions of Einstein's equations,''
  J.\ Math.\ Phys.\  {\bf 12}, 918 (1971).

\bibitem{Geroch:1972yt} 
  R.~P.~Geroch,
  ``A Method for generating new solutions of Einstein's equation. 2,''
  J.\ Math.\ Phys.\  {\bf 13}, 394 (1972).

\bibitem{VanProeyen:2001wr} 
  A.~Van Proeyen,
  ``Special geometries, from real to quaternionic,''
  hep-th/0110263.

 

\end{thebibliography}
\end{document}